\documentclass[12pt]{article}
\usepackage[letterpaper,margin=1in]{geometry}
\usepackage{newtxtext,newtxmath}
\usepackage{amsmath}
\usepackage{graphicx}
\usepackage{subfig}
\usepackage{booktabs}
\usepackage{algorithm}
\usepackage{algorithmic}
\usepackage{tabularx}
\usepackage{url}
\usepackage{authblk}
\usepackage[section]{placeins}
\usepackage[colorlinks=true,allcolors=blue]{hyperref}
\newcommand{\dfsqd}{DF-SQD}
\newcommand{\method}{\dfsqd}
\newcommand{\mha}{\,\mathrm{m}E_{\mathrm h}}
\newcommand{\hartree}{\,E_{\mathrm h}}

\title{DF-SQD: Deterministic Fields for Sampling-Based Quantum Diagonalization}
\author[1]{Kushagra Agarwal\thanks{Kushagra worked on this project during his summer internship at IBM Research India.}}
\author[1]{Anupama Ray\thanks{Corresponding author: anupamar@in.ibm.com}}
\affil[1]{IBM Research India}
\date{}

\begin{document}
\maketitle

\begin{abstract}
Sampling-based quantum diagonalization method exploits Quantum-centric supercomputing platforms to sample bitstrings for Hamiltonian projection on a quantum computer, and then classically diagonalize the Hamiltonian to estimate the eigen values and eigen vectors. In current quantum devices an algorithm is useful when shallow quantum circuits with error mitigation support can discover better results while having either a proof of convergence or some method to explain trust in experiment. In this paper, we introduce \dfsqd{}, a hybrid algorithm that derives deterministic auxiliary-field circuits from selected double-factorization leaves of the two-electron tensor. The circuits propose occupation-number configurations, while selected configuration interaction evaluates the original active-space Hamiltonian and can recentre subsequent proposal rounds. On $\mathrm{N}_2$ (32 qubits; 6-31G basis) and a 40-qubit $[\mathrm{Fe}_2\mathrm{S}_2(\mathrm{SCH}_3)_4]^{2-}$ active-space Hamiltonian, we show that \dfsqd{} improves the energy obtained from sampled determinant spaces while using shallow number-preserving circuits in both simulator and hardware runs. For $\mathrm{N}_2$, \dfsqd{}  is 45x more accurate with a 11.23\% smaller subspace, and due to its ability to sample better bitstrings at lesser shots it is 2.93x faster than SQD in quantum devices. 
For the iron--sulfur hardware data, \dfsqd{}   generated a subspace dimension of 221M with 400K shots, while SQD needed 1.5M shots to generate a 238M subspace, thus we have better subspace recovery evident from the hardware at 3.75x reduced shots. 
At a matched 50M subspace dimension , \dfsqd{} is 1.32x more accurate (achieves a 24.5\% relative error reduction over standard SQD). So overall, our method is able to discover better results with shallower circuits, is sample efficient, uses configuration recovery (so has targeted error mitigation) and we have empirical convergence observation. 

\end{abstract}

\section{Introduction}
Quantum chemistry is a prime candidate for quantum algorithms to be able to show advantage in the near-term. The most common task in theoretical quantum chemistry is the computation of ground-state energies by solving the Schrodinger equation in the Born-Oppenheimer approximation. Exact numerical solution in full configuration interaction grows combinatorially with the number of electrons and orbitals. There exists fault-tolerant quantum algorithms mostly based on quantum phase estimation (QPE) which could solve Schrodinger equation, however they require executing very deep circuits which is beyond the reach of current quantum devices. The Variational Quantum Eigensolver (VQE), proposed by Peruzzo et al.~\cite{peruzzo2014vqe}, was the first variational algorithm demonstrated for near-term devices; however, VQE faces severe scalability challenges due to measurement overhead~\cite{gonthier2022measurements} and difficult or prolonged optimization~\cite{fedorov2022vqe}.

In the quantum-utility regime~\cite{motta2024subspace}, researchers have proposed quantum extensions of classical SCI methods, including QSCI~\cite{kanno2026qsci}, and sampling-based quantum diagonalization (SQD) methods, which enabled chemistry experiments with up to 72 spin orbitals using up to 77 qubits~\cite{robledo2025sqd}. A heterogeneous quantum--classical fragmentation workflow has subsequently applied quantum configuration sampling on fragments of up to 94 qubits to protein--ligand complexes containing 11,608 and 12,635 atoms, with fragment solution and system-level assembly performed on classical supercomputers~\cite{merz2026crossing}. However these SQD methods rely on heavy classical post-processing and humongous classical resources and do not have convergence guarantees. Recently, the sample-based Krylov quantum diagonalization (SKQD) algorithm~\cite{yu2025skqd} was proposed with convergence guarantees under Krylov-diagonalization and ground-state-sparsity assumptions; quantum Krylov states are used as circuits from which samples are collected. However this algorithm suffers from very high depth circuits as depths of time-evolution circuits needed to generate Krylov vectors rapidly increase for many-body Hamiltonians in chemistry.

Quantum-assisted configuration-interaction methods offer a complementary strategy: prepare correlated states on a quantum processor, measure them in the occupation basis and use the observed configurations to define a classical variational problem~\cite{kanno2026qsci,robledo2025sqd,motta2024subspace}. These methods avoid measuring every Hamiltonian term on quantum hardware, but introduces a different resource trade-off. Proposed circuits must be shallow enough for present devices, their finite samples must cover determinants that improve the projected energy, and the product space generated from sampled spin sectors must remain classically tractable. Real-time Krylov constructions provide a direct connection to Hamiltonian dynamics but can require deep many-body evolution. Conversely, shallow ansatz-based sampling can be practical without supplying a general guarantee that physically important determinants receive appreciable probability~\cite{motta2024subspace}.

We address this trade-off with \dfsqd, a deterministic-field variant of SQD. Double factorization identifies one-body directions derived from the two-electron tensor~\cite{motta2021lowrank}. A Hubbard--Stratonovich construction~\cite{hubbard1959partition,hirsch1983discrete} turns ordinary-square factors of an auxiliary proposal generator into one-body orbital rotations, and a Hadamard sign design distributes a fixed circuit budget across deterministic field directions. Correlated pair seeds initialize the proposal circuits. The quantum processor therefore proposes determinant support rather than estimating energy: all Hamiltonian matrix elements and reported uncorrected energies are obtained from the original active-space tensors in a classical Rayleigh--Ritz solve.

The distinction between proposal generation and physical propagation is central. The circuit generator uses selected leaves, absolute factorization eigenvalues and ordinary operator squares, and separate circuit outcomes are pooled as a classical probability mixture. Its coherent average admits a short-time auxiliary-filter interpretation, but the implemented protocol is neither a physical-Hamiltonian Krylov construction nor quantum imaginary-time evolution~\cite{motta2020qite}. The applicable guarantees are operational: a fixed finite target set with nonzero mixture probability is recovered with increasing shots, and energies are monotone in explicitly nested cumulative uncapped determinant spaces.

We validate the algorithm first on N$_2$ in 6-31G basis and then on the $(30e,20o)$ singlet active-space Hamiltonian of $[\mathrm{Fe}_2\mathrm{S}_2(\mathrm{SCH}_3)_4]^{2-}$, a 40-qubit test of determinant discovery on simulator and hardware. Iron--sulfur clusters are established strongly correlated benchmarks for many-electron methods~\cite{sharma2014fes}. We further introduce an Epstein--Nesbet (EN)-inspired marginal-sector score tailored to the Cartesian spin-sector spaces used here and evaluate a deterministic, pool-restricted, denominator-clipped second-order correction as a secondary non-variational analysis~\cite{epstein1926stark,nesbet1955ci,sharma2017shci}.

\begin{figure}[h]
    \centering
    \includegraphics[width=1\linewidth]{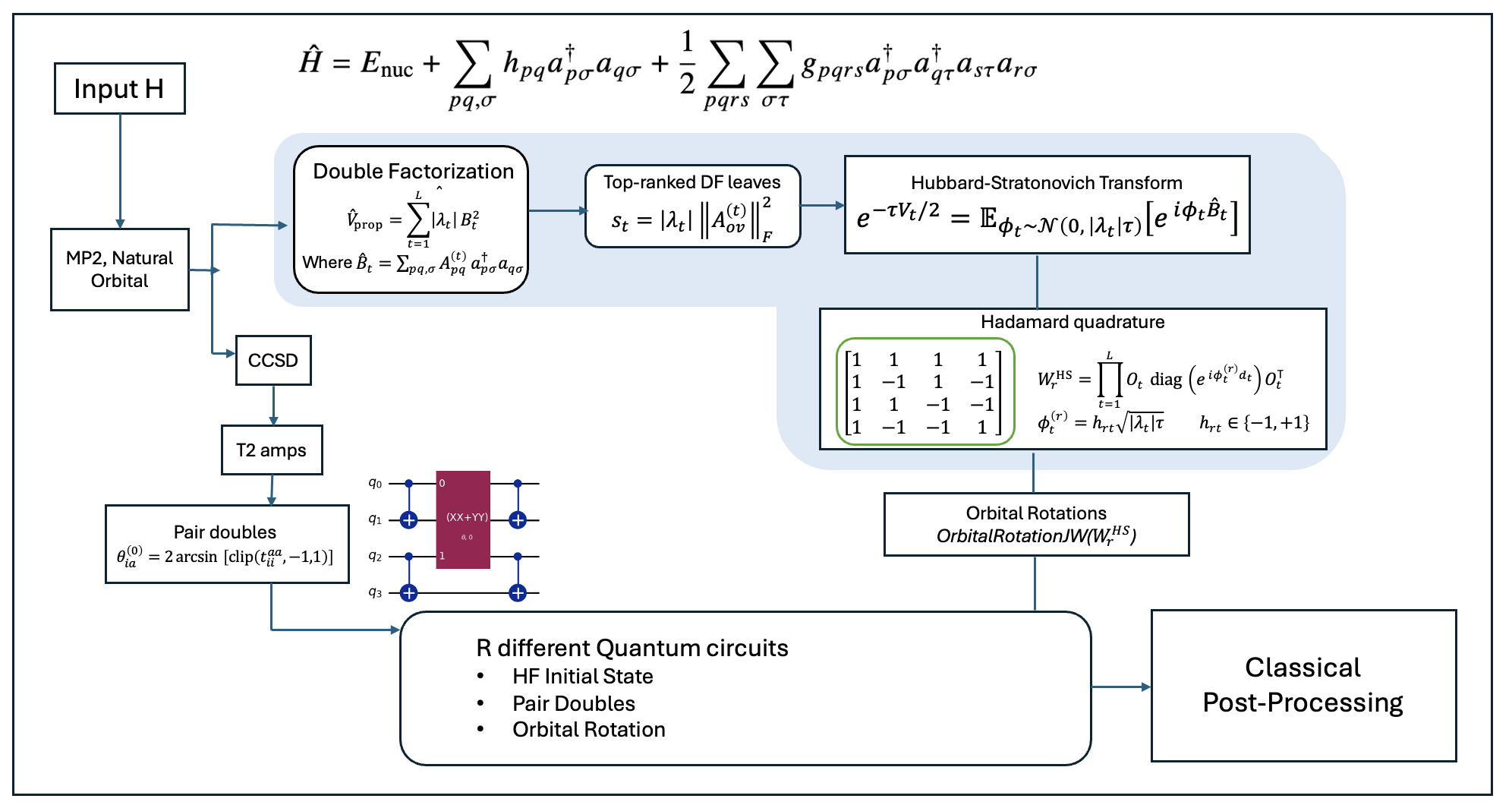}
    \caption{\textbf{Schematic overview of the \dfsqd{} algorithm.}
    The quantum stage combines CCSD-derived pair-double excitations with orbital-rotation proposals constructed from double factorization, the Hubbard--Stratonovich transformation, and deterministic Hadamard-sign field schedules. Classical post-processing consists of configuration recovery, spin completion, particle-number postselection, top-$K$ spin-sector selection, selected-CI, and an optional second-order perturbative (PT2) correction.}
    \label{fig:algorithm-overview}
\end{figure}

\section{Algorithm overview}
\dfsqd{} separates quantum support generation from classical energy evaluation. Starting from an active-space Hamiltonian and fixed alpha- and beta-electron numbers, correlated pair seeds and Hamiltonian-derived orbital rotations generate occupation-number samples. Strict particle-number postselection, with configuration recovery for noisy hardware data, produces candidate alpha and beta strings. A classical selected-CI solve then evaluates the physical Hamiltonian in a retained Cartesian product of those strings. The general implementation can use selected-CI amplitudes to update later proposal rounds, but the reported N$_2$ and Fe$_2$S$_2$ results use only the initial \texttt{flip0}\footnote{\texttt{flip0} denotes the initial, pre-feedback circuit ensemble; \texttt{flip$n$} denotes the cumulative sample pool after adaptive round $n$. The label refers to an ensemble/checkpoint, not to a physical spin flip.} circuit ensemble and therefore do not test adaptive feedback on hardware. Figure~\ref{fig:algorithm-overview} summarizes the workflow; mathematical construction and implementation details are given in Methods and Supplementary Notes S1--S7.

\section{Results}
In this section we present results on two molecules $\mathrm{N}_2$ and $[\mathrm{Fe}_2\mathrm{S}_2]$. For Nitrogen, at equilibrium bond length, we used 6-31G basis with (10e,16o) active space, resulting in 32 qubits. For $[\mathrm{Fe}_2\mathrm{S}_2]$, at equilibrium bond length, we used TZP-DKH basis with (30e,20o) active space resulting in 40qubits. Below we present detailed results on both simulator and real hardware. The ablation studies have been moved to Supplementary material along with details of proposed method.

\subsection{N$_2$ results}

\begin{figure}[H]
\centering
\subfloat[Energy error.]{\includegraphics[width=0.47\textwidth]{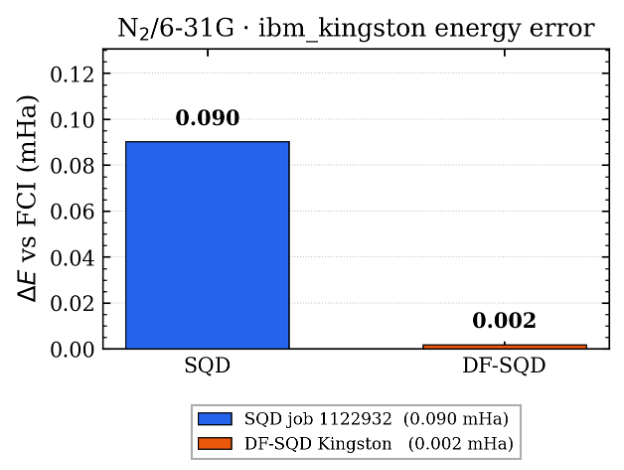}}\hfill
\subfloat[Selected product-space dimension.]{\includegraphics[width=0.47\textwidth]{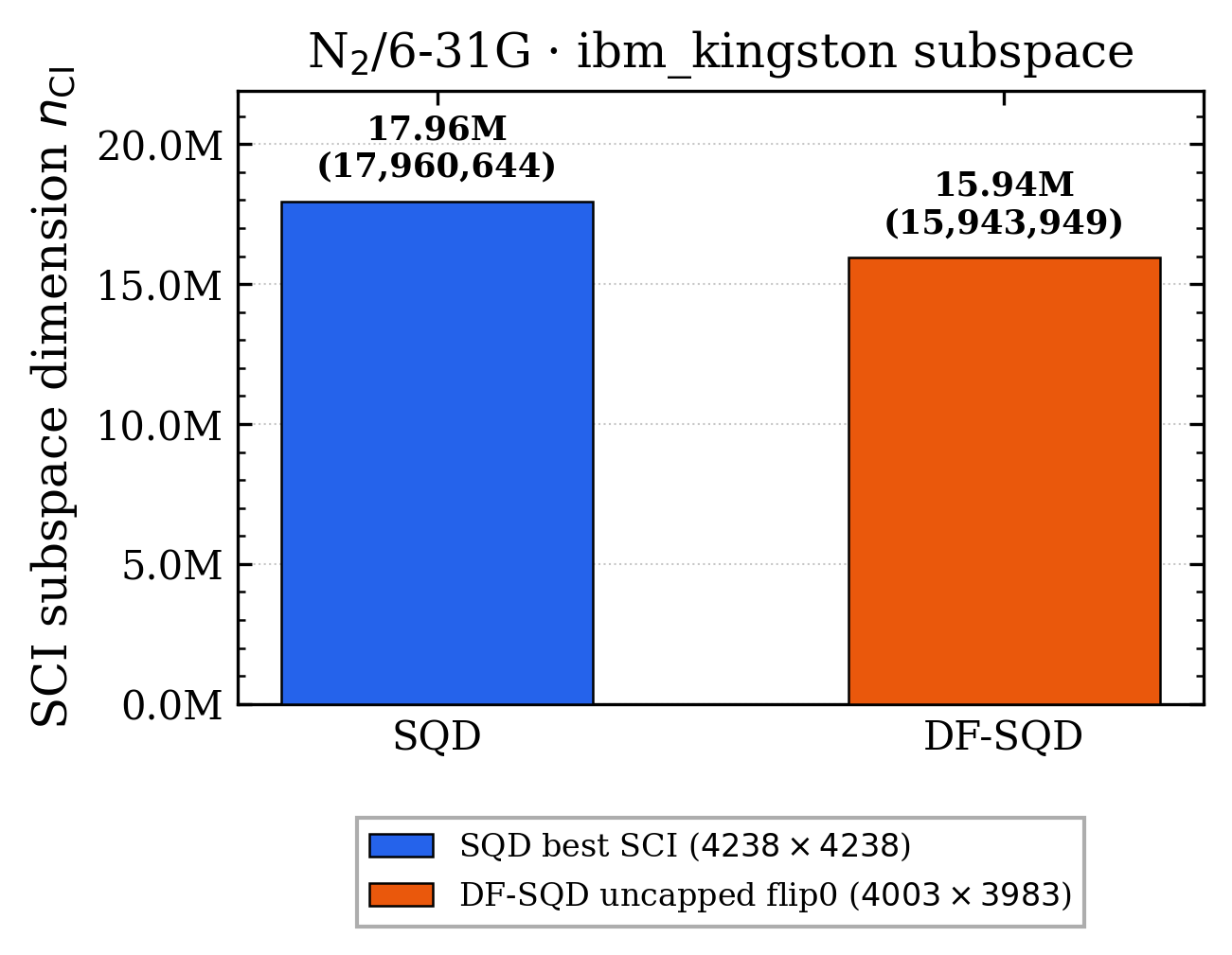}}\\
\caption{\textbf{N$_2$ validates determinant support generated on IBM Quantum hardware.} (a) Error relative to the CASSCF/FCI reference of $-109.10301625689416\hartree$. (b) Dimension of the alpha--beta determinant product used for the final projected solve. The compared methods use different circuit ensembles and shot allocations.}
\label{fig:n2-hardware}
\end{figure}

We first tested \method{} on N$_2$ at an internuclear separation of $1.0977$~\AA{} in the 6-31G basis. The benchmark uses a $(16o,10e)$ active space and a CASSCF/FCI reference of $-109.10301625689416\hartree$. The comparator is SQD with a local-unitary cluster-Jastrow (LUCJ) proposal circuit~\cite{motta2023lucj,robledo2025sqd}. The hardware data were obtained on \texttt{ibm\_kingston}; the \method{} run used 12 circuits with 8,192 shots per circuit. At the equilibtrium bond length, the projected errors were approximately $0.090\mha$ for the SQD comparator and $0.00188\mha$ for \method, with product spaces of approximately $17.96$ million and $15.94$ million determinants, respectively (Fig.~\ref{fig:n2-hardware}a,b).

Because \method{} distributes its samples over a circuit ensemble whereas the comparator uses a different circuit and shot allocation, Table~\ref{tab:n2_resources} reports the measured quantities for 2q-depth, sampling time, and QPU usage without attributing any difference to one algorithmic component.

\begin{table}[H]
\centering
\small
\begin{tabularx}{\linewidth}{@{}XcXX@{}}
\toprule
\textbf{Metric} &
\textbf{SQD (LUCJ)} &
\textbf{\method{} (12 fields)} &
\textbf{Improvement} \\
\midrule
FakeKingston 2-qubit depth &
137 &
104 (82--118) &
24\% lower \\

IBM Kingston 2-qubit depth &
137 &
mean 121.25; median 119.5; range 101--151 &
run-specific \\

Aer MPS sampling time &
21.3 min &
6.1 min &
$3.5\times$ faster \\

IBM Quantum QPU usage &
85 s &
29 s &
$2.9\times$ lower \\
\bottomrule
\end{tabularx}

\caption{\textbf{N$_2$/6-31G resource accounting.} Depth entries are transpiled two-qubit depth; parenthetical values give ranges unless otherwise stated. Simulator depths used multi-seed SABRE layout selection with 16 seeds, whereas the hardware circuits were transpiled at optimization level 3 without multi-seed SABRE. The reported hardware depths are therefore those of the executed circuits; applying the same multi-seed layout search would be expected to reduce them. SQD-LUCJ used 300,000 shots, whereas \method{} used 98,304 shots (12 circuits of 8,192 shots). Sampling wall time and QPU usage correspond to these distinct allocations and are run-specific resource accounting, not a controlled algorithmic comparison.}

\label{tab:n2_resources}
\end{table}

\begin{figure}[t]
\centering
\subfloat[Baseline energy error.]{\includegraphics[width=0.47\textwidth]{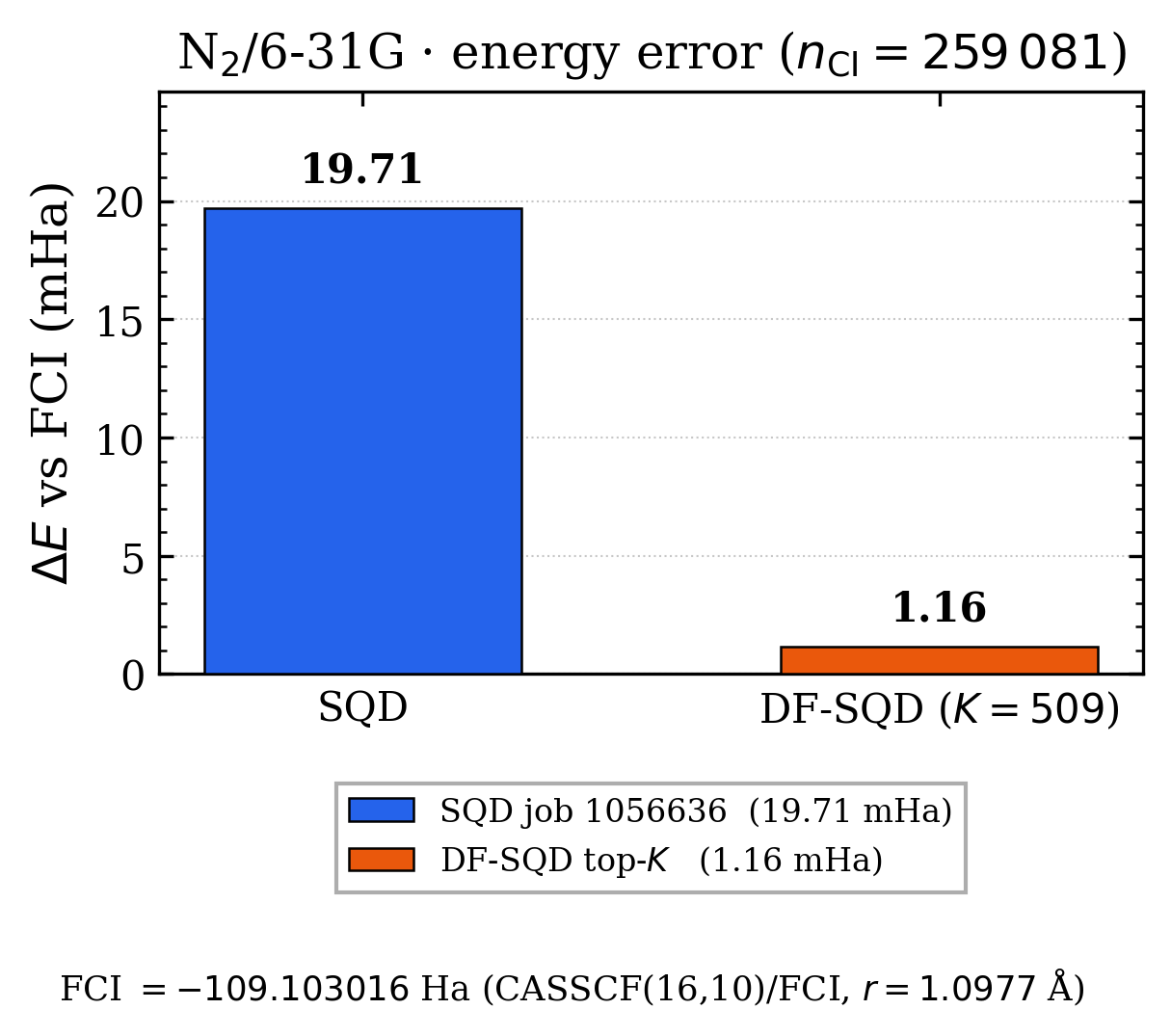}}\hfill
\subfloat[Baseline product-space dimension.]{\includegraphics[width=0.47\textwidth]{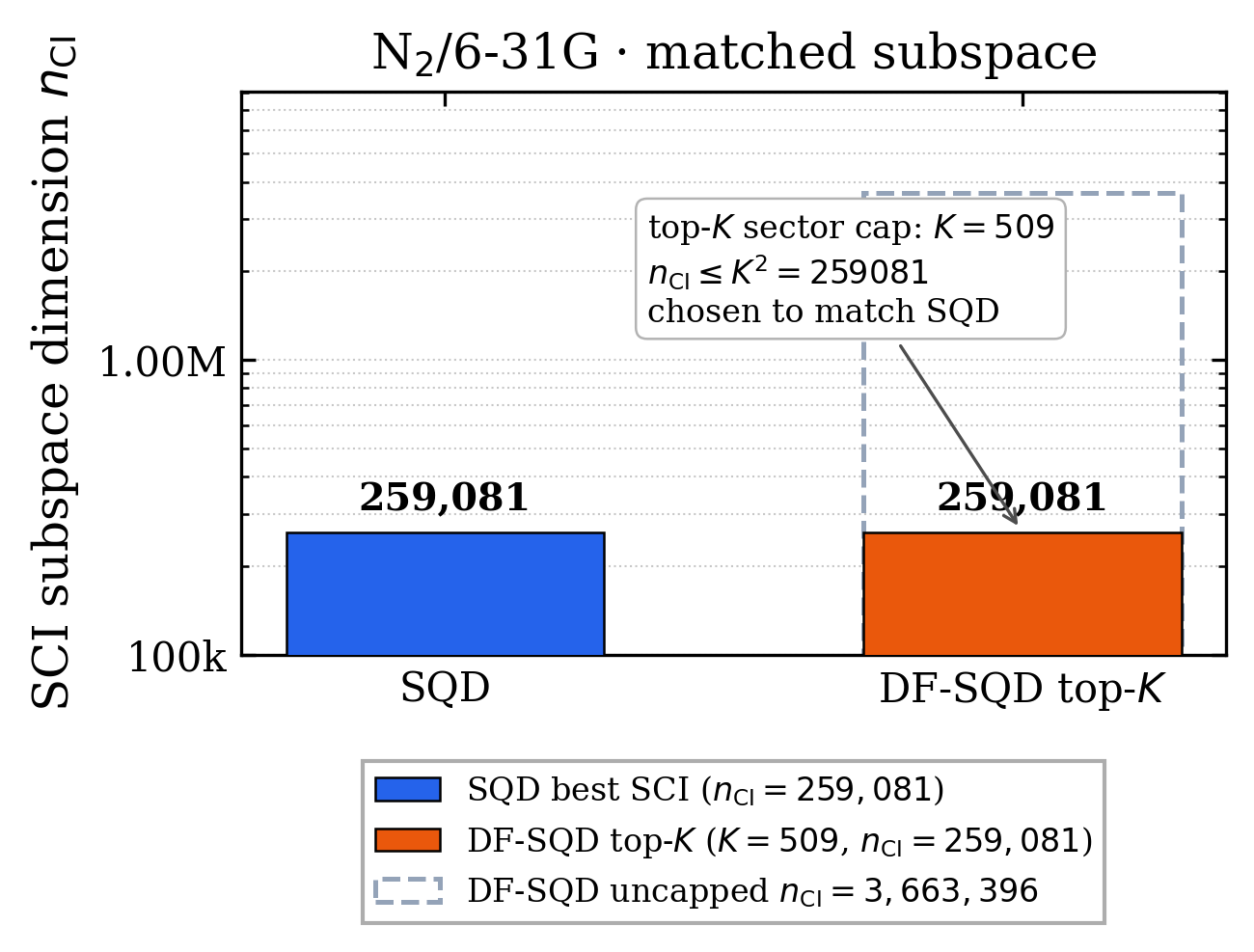}}\\
\caption{\textbf{N$_2$ simulator and baseline controls.} The panels report the corresponding energy error and selected product-space dimension for the N$_2$ baseline study.}
\label{fig:n2-baseline}
\end{figure}

The corresponding simulator and baseline controls are shown in Fig.~\ref{fig:n2-baseline}. These panels provide the energy and selected-space context for the hardware comparison.


\subsection{Fe$_2$S$_2$ simulation results}

We next considered the $M_S=0$ sector of the $[\mathrm{Fe}_2\mathrm{S}_2(\mathrm{SCH}_3)_4]^{2-}$ Hamiltonian in a $(30e,20o)$ active space, corresponding to 40 Jordan--Wigner qubits with $N_\alpha=N_\beta=15$, and compared it with a singlet reference~\cite{jordan1928wigner,sharma2014fes}. All reported errors use the Li--Chan bond-dimension-$8{,}000$ density-matrix-renormalization-group (DMRG) value $E_{\mathrm{ref}}=-116.6056091\hartree$ for this active-space Hamiltonian~\cite{chan2011dmrg,li2017spinprojected,li_chan_fes_repository}. The simulator circuits were compiled to the FakeKingston target and sampled with a matrix-product-state simulator of maximum bond dimension 256.

\begin{figure}[h]
\centering
\subfloat[Energy error.]{\includegraphics[width=0.47\textwidth]{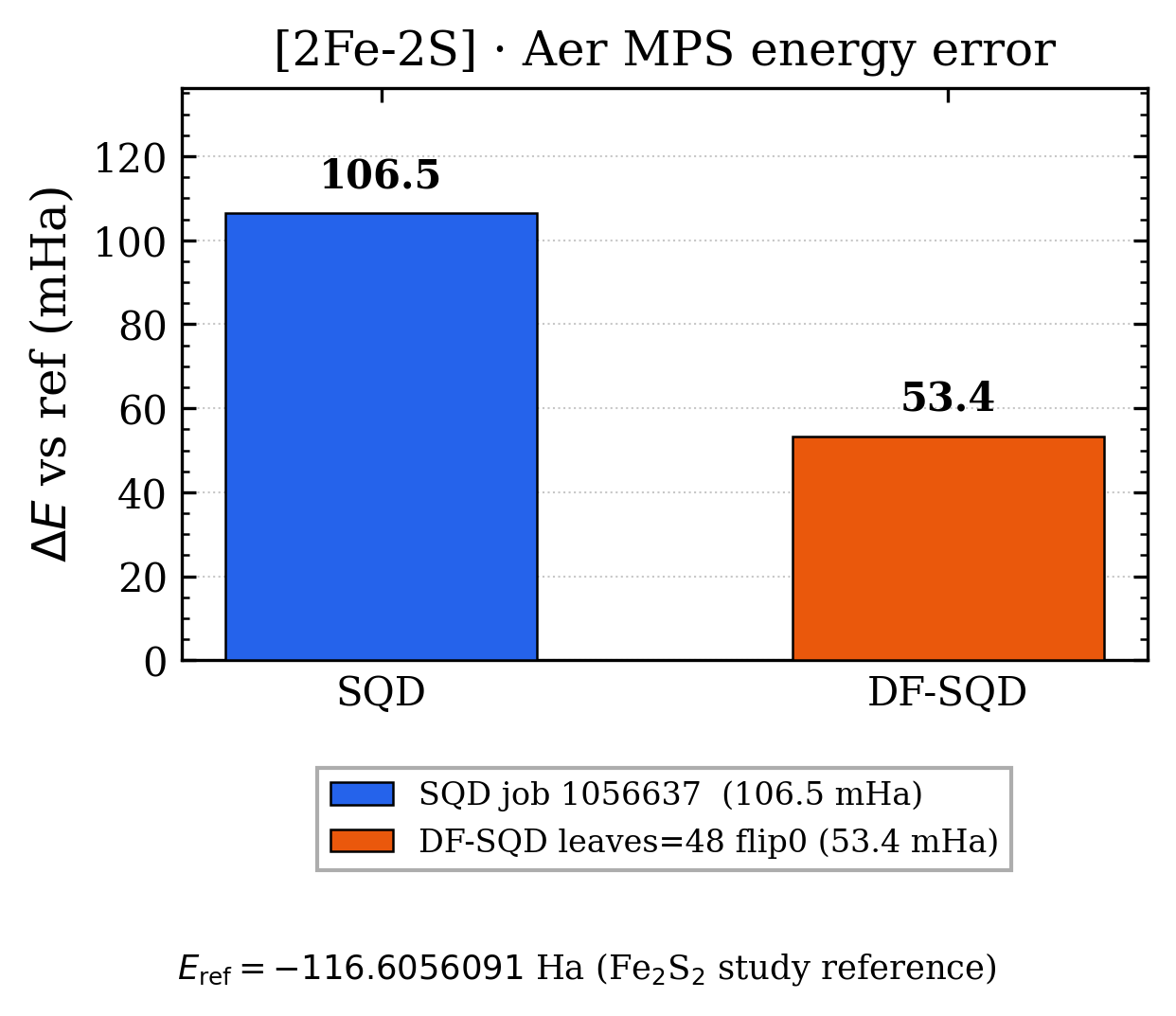}}\hfill
\subfloat[Selected product-space dimension.]{\includegraphics[width=0.47\textwidth]{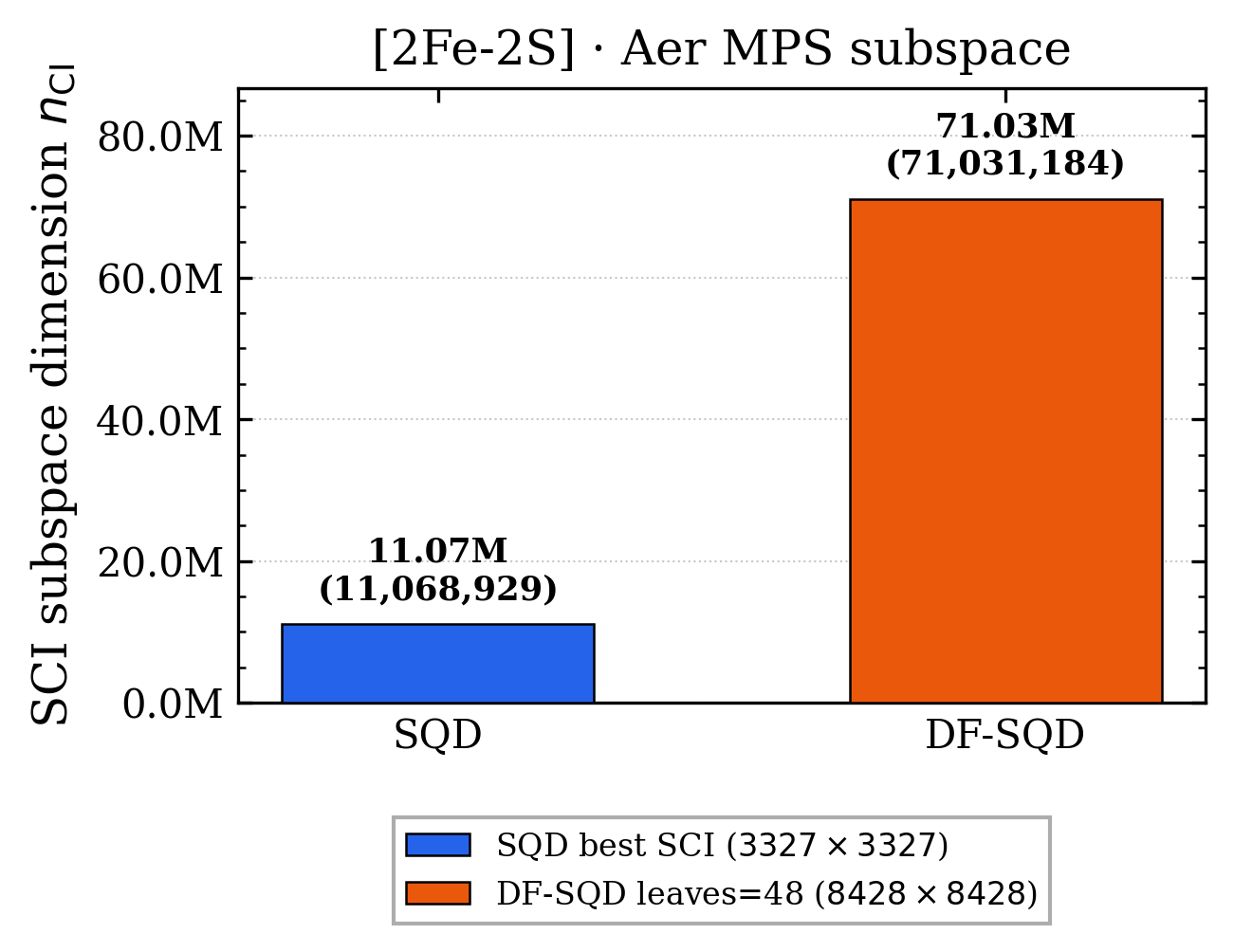}}
\caption{\textbf{DF-SQD constructs a richer Fe$_2$S$_2$ determinant space in simulation.} The $(30e,20o)$ active-space Hamiltonian is evaluated relative to $-116.6056091\hartree$. (a) Variational energy error. (b) Dimension of the selected alpha--beta product space. In (Fig.~\ref{fig:fe2s2-hardware}a,b) we show \method{} gives lower errors when subspace is capped to 50M for both algorithms on hardware.}
\label{fig:fe2s2-sim}
\end{figure}

In the simulator comparison, \method{} reduced the error from approximately $106.5$ to $53.4\mha$ (Fig.~\ref{fig:fe2s2-sim}a), while increasing the selected product space from approximately $11.1$ million to $71.0$ million determinants (Fig.~\ref{fig:fe2s2-sim}b). The result therefore demonstrates richer determinant discovery, not an accuracy gain at a matched classical dimension. Table~\ref{tab:fe2s2-sim-resources} records the associated depth, shot and simulator-sampling metrics; the timings apply to this simulator configuration and are not asymptotic runtime estimates.

\begin{table}[H]
\centering
\small
\begin{tabularx}{\linewidth}{@{}XcX@{}}
\toprule
\textbf{Metric} & \textbf{SQD-LUCJ} & \textbf{\method{}} \\
\midrule
Transpiled two-qubit depth & 183 & mean 173.92; median 172.5; range 145--193 \\
Total shots & 1,500,000 & 196,608 (24 $\times$ 8,192) \\
Aer MPS sampling & 4,905.743973 s (81.76 min) & 1,810.011425 s (30.17 min) \\
\bottomrule
\end{tabularx}
\caption{\textbf{Fe$_2$S$_2$ simulator resource metrics.} Values describe the recorded FakeKingston/Aer MPS runs with different circuit ensembles and shot allocations; they are run-specific accounting rather than a controlled speedup comparison.}
\label{tab:fe2s2-sim-resources}
\end{table}

\subsection{Hardware sampling improves Fe$_2$S$_2$ accuracy at matched classical budget}

\begin{figure}[h]
\centering
\subfloat[Matched-budget energy error.]{\includegraphics[width=0.47\textwidth]{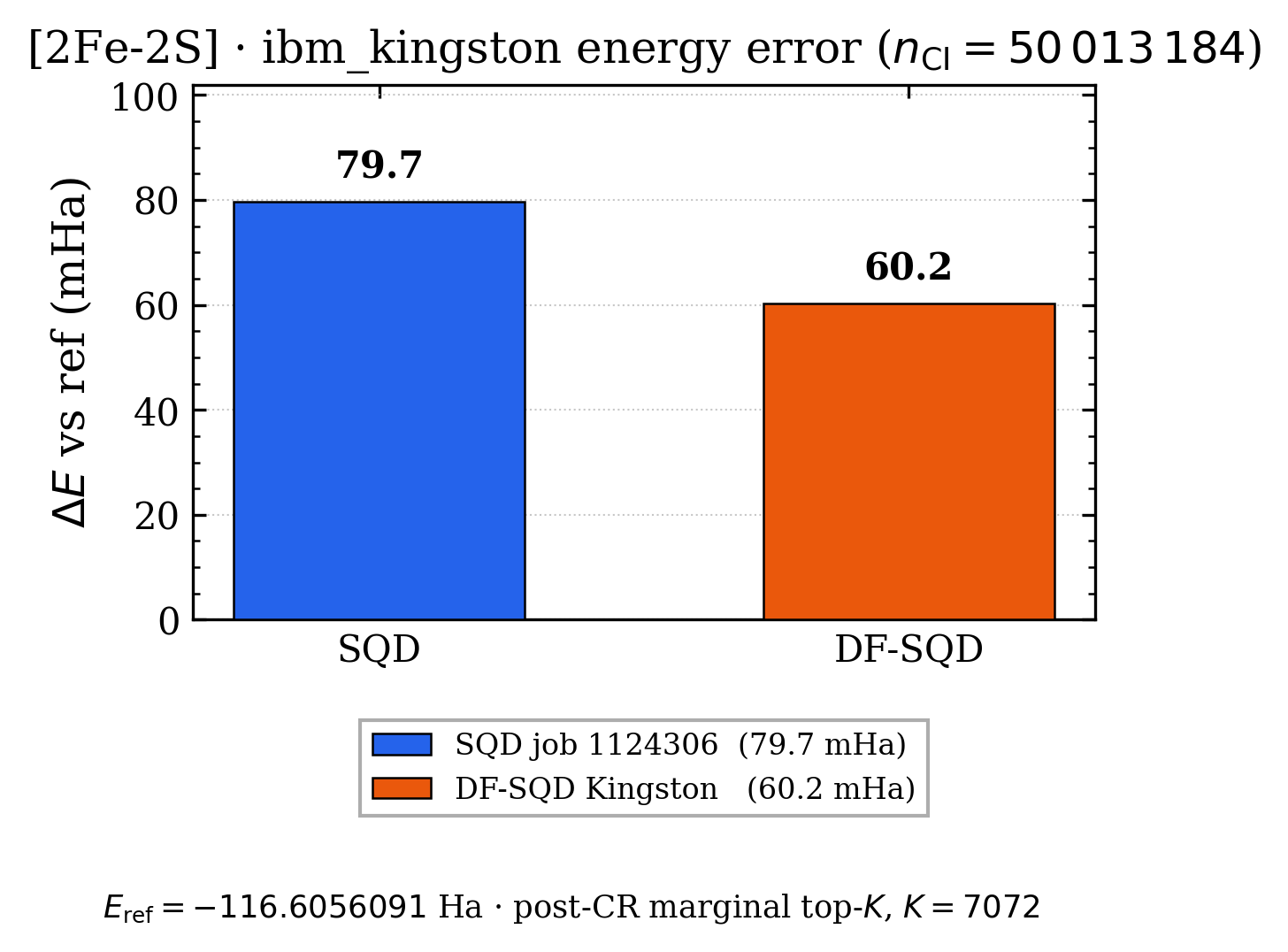}}\hfill
\subfloat[Available and selected spaces.]{\includegraphics[width=0.47\textwidth]{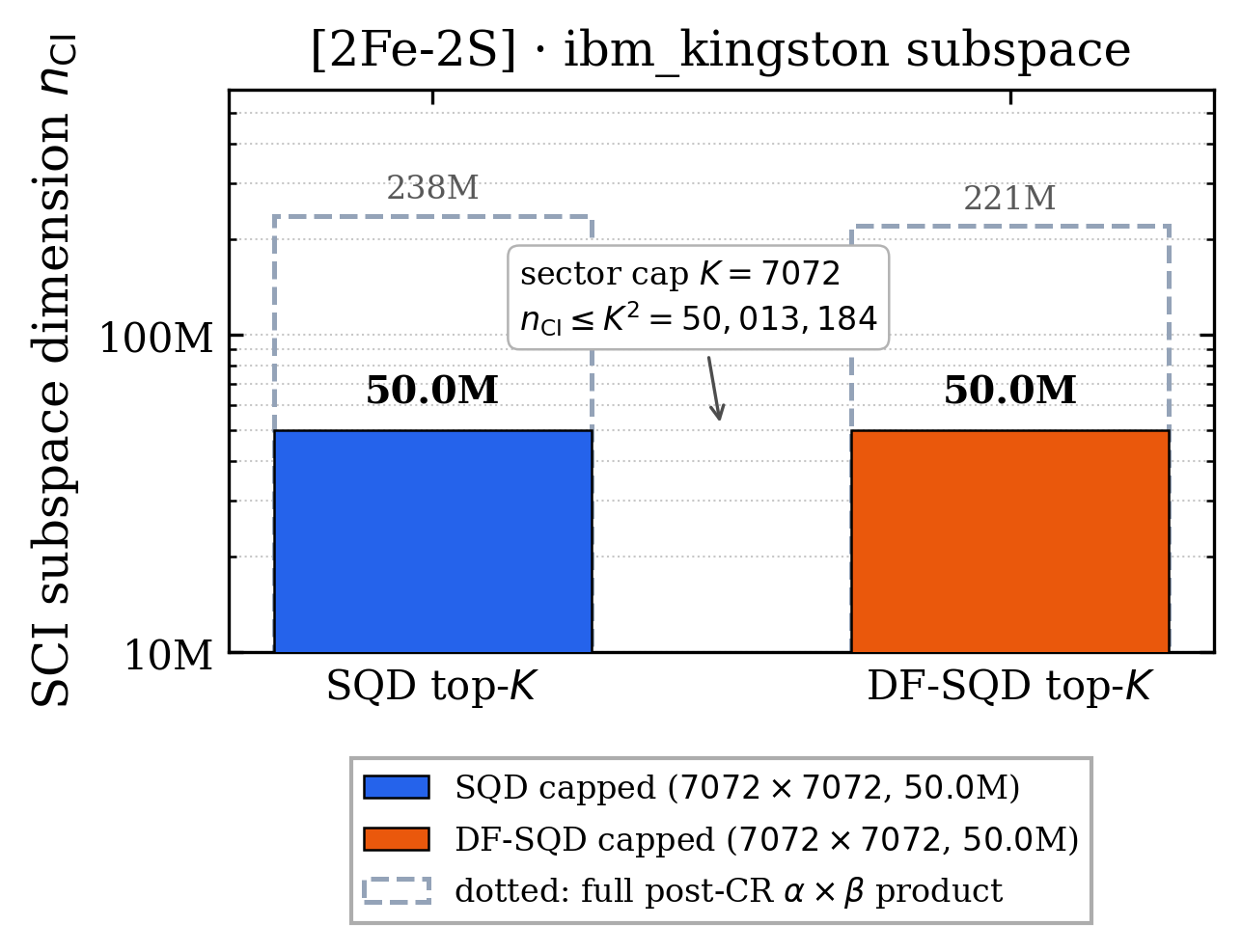}}
\caption{\textbf{DF-SQD improves the Fe$_2$S$_2$ variational energy at a matched 50-million-determinant budget.} (a) Error of the selected-CI energy relative to the Li--Chan DMRG reference. (b) Full post-recovery Cartesian products and matched $K=7,072$ sector products. }
\label{fig:fe2s2-hardware}
\end{figure}

The Fe$_2$S$_2$ hardware experiment compares the post-recovery SQD-LUCJ and \method{} sample pools. The SQD-LUCJ data used one circuit with 1,500,000 shots. The \method{} data used 24 auxiliary-field circuits with 16,384 shots each, for 393,216 total shots, followed by spin-flip augmentation. Both pools were reduced to $K=7,072$ strings in each spin sector, giving the same nominal product-space dimension, $K^2=50,013,184$. At this matched classical budget, the variational error was $79.7\mha$ for SQD-LUCJ and $60.2\mha$ for \method{} (Fig.~\ref{fig:fe2s2-hardware}a,b). The full post-configuration-recovery products shown in the subspace panel are available Cartesian spaces before the matched sector cap, not spaces diagonalized for this comparison. Table~\ref{tab:fe2s2-hardware-resources} records circuit, shot, depth and QPU usage metrics. Because shot totals, circuit counts and batching differ, these measurements describe the resource profile of the reported runs rather than a hardware-independent speedup. Configuration recovery is an occupancy-guided heuristic followed by strict $(15,15)$ particle-number postselection~\cite{robledo2025sqd}.
\begin{table}[h]
\centering
\small
\begin{tabularx}{\linewidth}{@{}XcX@{}}
\toprule
\textbf{Metric} & \textbf{SQD-LUCJ} & \textbf{\method{}} \\
\midrule
Circuits & 1 & 24 \\
Shots per circuit & 1,500,000 & 16,384 \\
Total shots & 1,500,000 & 393,216 \\
Transpiled two-qubit depth & 183 & mean 173.92; median 172.5; range 145--193 \\
QPU usage & 426 s & 112 s \\
\bottomrule
\end{tabularx}
\caption{\textbf{Fe$_2$S$_2$ hardware resource metrics.} Values are run-specific accounting for the reported \texttt{ibm\_kingston} jobs, which used different circuit ensembles, shot totals and batching; no hardware-independent speedup is inferred.}
\label{tab:fe2s2-hardware-resources}
\end{table}

\subsection{Hamiltonian-aware marginal-sector selection improves sampled support}

Marginal frequency is a transparent way to cap the classical space, but it need not identify the strings that couple most strongly to the current selected-CI state. We therefore evaluated a custom EN-inspired marginal-sector score on the finite post-recovery product pool. The procedure computes determinant-pair scores from one $Hc$ contraction, sums those scores separately over the opposite spin sector and retains the top $K$ alpha and beta strings. It draws on conventional EN and selected-CI ranking ideas ~\cite{epstein1926stark,nesbet1955ci,huron1973cipsi,sharma2017shci}, but it is not determinant-level CIPSI: its selected space remains a Cartesian sector product.

\begin{figure}[h]
\centering
\subfloat[EN-inspired variational and clipped second-order $K$-sweeps.]{\includegraphics[width=0.50\textwidth]{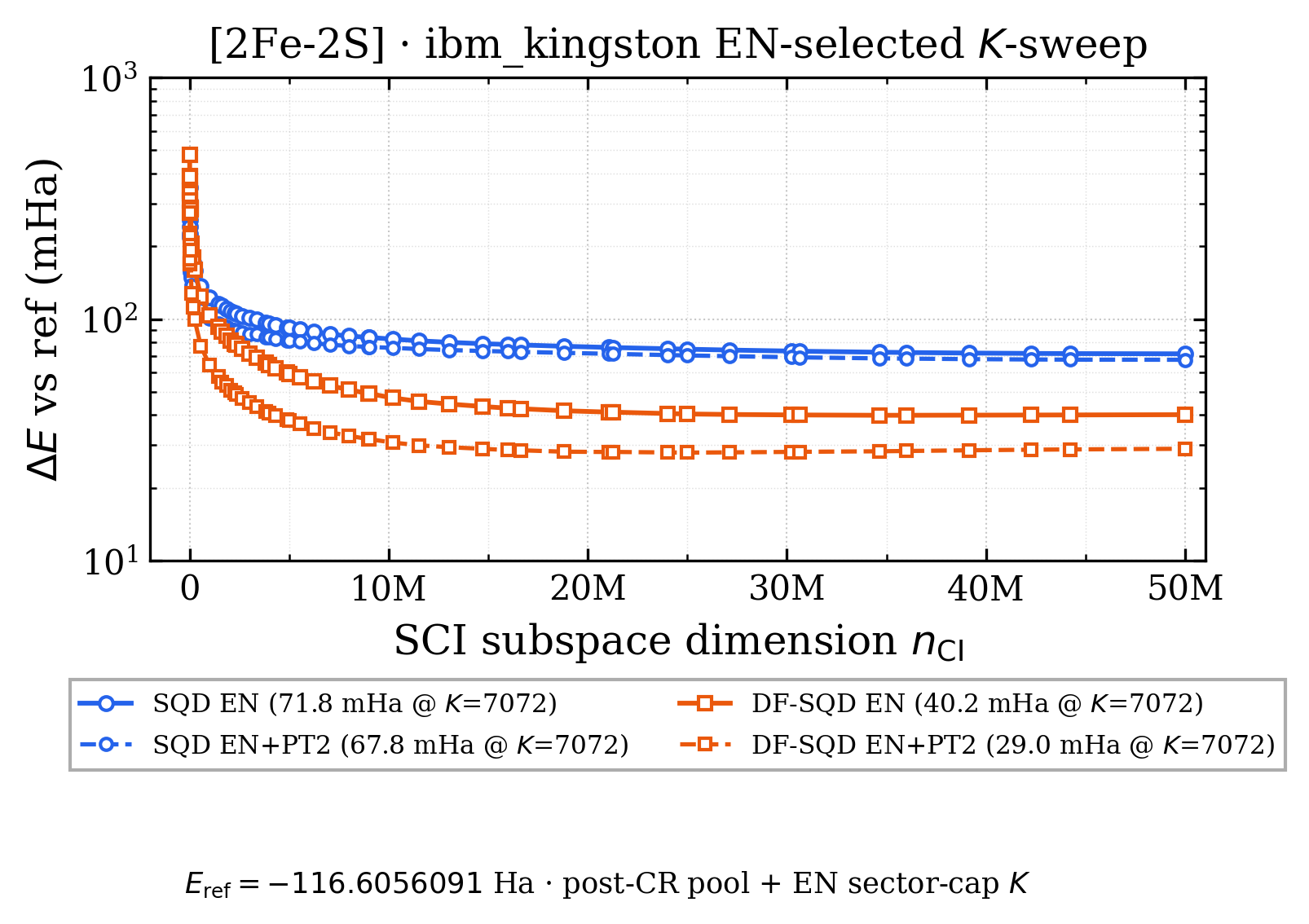}}\hfill
\subfloat[Selection and PT2 contributions for DF-SQD.]{\includegraphics[width=0.45\textwidth]{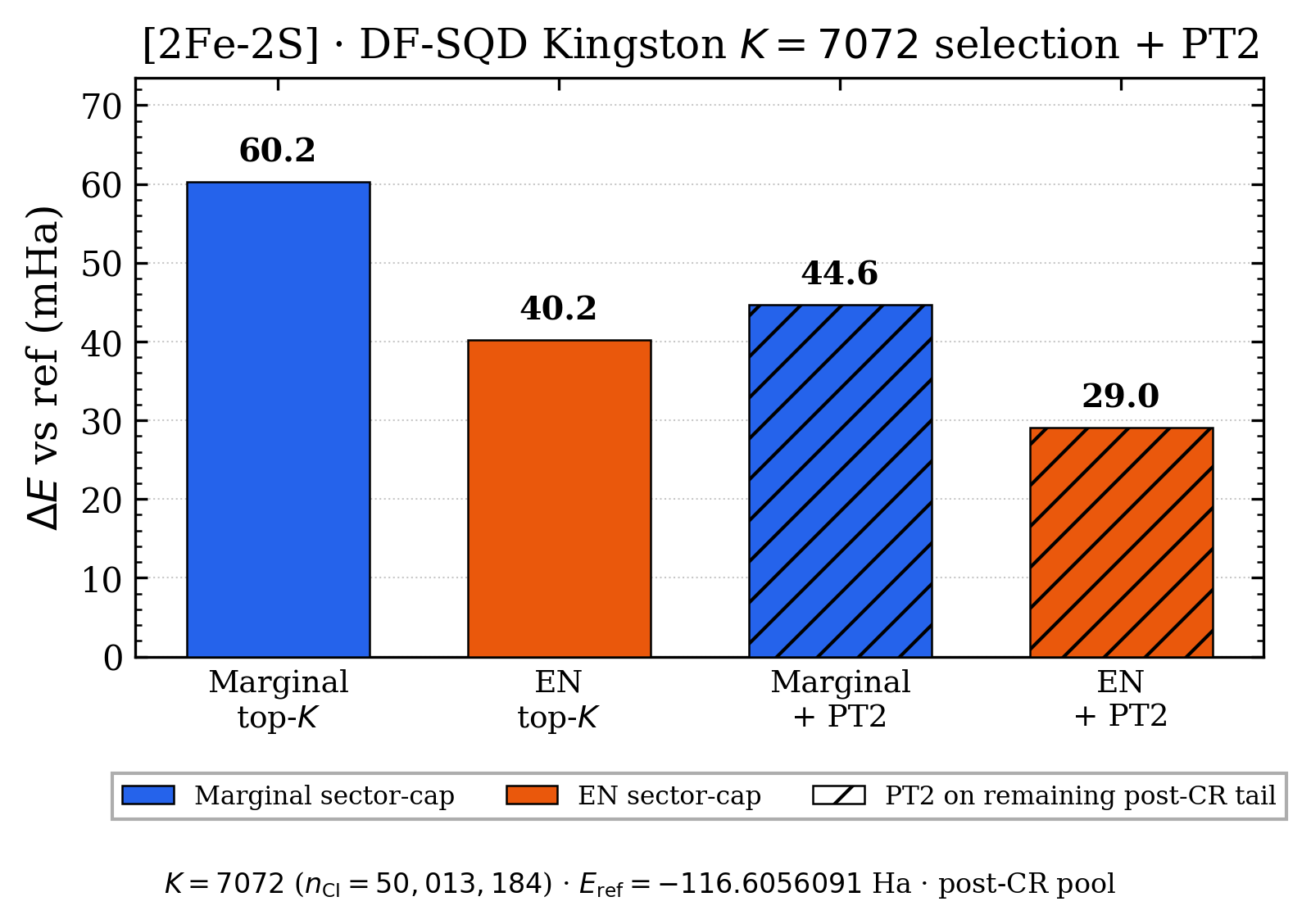}}
\caption{\textbf{Hamiltonian-aware marginal-sector selection extracts additional value from the sampled Fe$_2$S$_2$ pool.} (a) Approximate selected-CI variational errors and pool-restricted, denominator-clipped second-order estimates as functions of Cartesian product-space dimension. Each reverse sweep uses nested prefixes of a single alpha/beta ranking computed from the $K=7,072$ reference state; it is not an independently iterated CIPSI calculation at every $K$. (b) Decomposition of the error reduction from frequency selection, EN-inspired marginal-sector selection and the post hoc PT2 estimate.}
\label{fig:en-pt2}
\end{figure}

For the reverse $K$-sweep, one score pass from the final $K=7,072$ state defined fixed alpha and beta rankings, and smaller spaces used nested prefixes without reranking. The selected-CI energies contain departures from the monotonic ordering expected of fully converged nested Rayleigh--Ritz solves; we therefore treat the sweep as an approximate iterative-solver diagnostic rather than evidence of variational monotonicity. At $K=7,072$, EN-inspired marginal-sector selection reduced the \method{} variational error from the frequency-selected $60.2\mha$ to $40.2\mha$ (Fig.~\ref{fig:en-pt2}). The corresponding SQD-LUCJ EN-selected error was $71.8\mha$.

We additionally evaluated a deterministic EN second-order correction over the remainder of each finite post-recovery sector product. This pool-restricted EN-PT2 is a post hoc estimate: it neither changes the quantum samples nor enters adaptive feedback, and $E_{\mathrm{var}}+E^{(2)}$ is not covered by the variational upper-bound claim. At $K=7,072$, it changed the \method{} error from $40.2$ to $29.0\mha$ and the SQD-LUCJ error from $71.8$ to $67.8\mha$. This secondary analysis shows nonzero Hamiltonian coupling to determinants retained in the recovered pool beyond the selected 50-million-determinant space; it does not provide an independent convergence guarantee.


\subsection{Conditional support and accessible-space bounds}

The executed circuits define a classical mixture $q(D)$ over determinants. For any fixed finite target set $\mathcal T$ satisfying $q(D)\ge q_{\min}>0$, independent sampling gives a failure probability no larger than $|\mathcal T|e^{-Nq_{\min}}$ after $N$ mixture draws. This result describes recovery of support already assigned nonzero circuit probability; it does not establish that all physically important determinants satisfy a useful lower bound.

When all observed alpha and beta strings are retained cumulatively, their Cartesian product spaces are nested. Rayleigh--Ritz energies in those uncapped spaces are therefore nonincreasing and converge to the minimum energy representable by the circuit-accessible product space. Exact active-space convergence follows only if that accessible space contains the exact ground-state wavefunction. These statements do not apply to every capped feedback solve, and they are independent of the limited coherent-filter interpretation of the auxiliary-field construction.

\section{Discussion}
\dfsqd{} reframes the role of the quantum processor in subspace diagonalization. Rather than approximating an energy or reproducing physical time evolution, its circuits generate structured determinant proposals derived from the interaction tensor. The original Hamiltonian then determines which proposals are useful. This separation permits shallow, number-preserving circuits while preserving a conventional variational interpretation for the final uncorrected selected-CI energy.

The experiments identify three possible sources of improvement: double-factorized auxiliary-field rotations broaden sampled support along Hamiltonian-derived one-body directions; deterministic field signs make the selected directions reproducible; and classical selected-CI or EN-inspired marginal-sector ranking allocates a fixed product-space budget using Hamiltonian information. The present data do not isolate these effects completely, so the observed gains should be attributed to the complete workflow rather than to the Hadamard design alone.

The method deliberately exchanges quantum depth for classical subspace cost. Keeping $K$ alpha and $K$ beta strings produces as many as $K^2$ determinants, and the uncapped Cartesian product can be substantially larger. The matched-$K$ iron--sulfur comparison and the EN analysis therefore report accuracy together with selected-space dimension. 

The scope of the formal results is specific. Finite-shot support recovery requires positive probability under a fixed executed circuit mixture; it does not prove that every determinant important to the exact ground state has useful probability. Rayleigh--Ritz monotonicity applies to nested cumulative uncapped spaces, not to every changing top-$K$ solve. Also configuration recovery is a heuristic method followed by strict particle-number postselection. We evaluated hardware-topology-based postselection in our ablation studies, but did not include it in the final algorithm. In the 24-circuit Fe$_2$S$_2$ postselection run (393,216 total shots), node-based postselection retained 18,033 shots (4.59\%) and edge-based postselection retained 184,021 shots (46.8\%). Despite reducing the sampled spaces, diagonalizing the postselected spaces produced larger energy errors than frequency-capped spaces of similar dimension constructed without topology-based postselection. We therefore retained occupancy-guided configuration recovery followed by strict particle-number postselection, as components of \dfsqd{}. To extract additional value from the portion of the recovered pool excluded by top-$K$ selection, we employ the denominator-clipped, pool-restricted second-order (PT2) estimates. However, these are post hoc and non-variational; they do not alter quantum sampling or the primary variational claims.

Alternative approaches to reducing quantum circuit depth, such as Approximate Quantum Compiling (AQC)~\cite{robertson2025approximate} and Auxiliary-Field Quantum Monte Carlo (AFQMC)~\cite{motta2018abinitio}, were considered but ultimately not adopted and is a future work for us. AQC methods frequently rely heavily on classical optimization to compress time-evolution unitaries, while AFQMC generates significant fractions of the core correlation energy directly from classical stochastic walkers guided by a trial state. Our focus was explicitly on shifting as much of the chemical representation as possible directly onto the quantum processor via shallow, physically inspired circuits, rather than relying on heavy classical surrogate optimization to yield the energy.

This alignment of algorithmic components---ensuring low depth, high chemical relevance on the quantum hardware, and inspiration from robust analytical methods---was significantly accelerated by an AI-guided (Large Language Model) discovery process. Rather than passively generating code, we used the LLM to review literature to help propose candidate design components; each of which was validated by isolated empirical tests before inclusion. While many of the AI-generated hypotheses got rejected for violating realistic implementation constraints or contradicting prior empirical observations, the remaining ideas were quantified and rigorously vetted via small, isolated empirical tests before being assembled into the final \dfsqd{} workflow.


These results motivate tests across more active spaces, backends and matched shot budgets. Particularly informative next steps include repeated hardware runs that quantify calibration and sampling variability. Within the stated limits, \dfsqd{} provides a practical route for using shallow circuits to guide large classical configuration-interaction calculations.

\section{Methods}
The general \method{} workflow is explained in  Algorithm \ref{alg:dfsqd}. \method{} separates determinant-support generation from evaluation of the physical Hamiltonian. In addition to the circuit and shot budgets, its parameters specify the deterministic field-time scale, correlated seed policy, recovery and spin-flip policies, adaptive feedback, and the final product-space rule.

\begin{algorithm}
\caption{General DF-SQD support-generation and variational-solve algorithm}
\label{alg:dfsqd}
\begin{algorithmic}[1]
\REQUIRE Tensors $(h,g,E_{\mathrm{nuc}})$; electron numbers $(N_\alpha,N_\beta)$; hyperparameters; policies; final-space rule $\mathcal F$
\ENSURE Variational energy $E_{\mathrm{var}}$
\STATE Rotate $(h,g)$ to MP2 natural orbitals and run CCSD to obtain proposal amplitudes
\STATE Factorize $g_{pqrs}=\sum_t\lambda_t A^{(t)}_{pq}A^{(t)}_{rs}$, rank leaves, retain top $L$, and set time steps $\tau_m$
\STATE Select top $P$ CCSD pair amplitudes and set initial seed angles $\theta_{ia}^{(0)}=2\arcsin[\operatorname{clip}(t_{ii}^{aa},-1,1)]$
\STATE Initialize cumulative count maps, correlated seeds, and Thouless rotation
\FOR{$a=0,\ldots,A$}
  \IF{$a>0$}
    \STATE Distill preceding selected-CI marginals into ranking-only pseudo-counts
    \STATE Extract selected-CI single and linked-double ratios; damp as $r^{(a)}=(1-\gamma)r^{(a-1)}+\gamma r^{\mathrm{SCI}}$
    \STATE Update Thouless angles $\kappa_i^a=\operatorname{clip}(\arctan r_i^a,-\kappa_{\max},\kappa_{\max})$ and seed angles $\theta=2\arctan r$
  \ENDIF
  \FOR{$r=0,\ldots,R-1$}
    \STATE Set fields $\phi_t^{(r)}=h_{rt}\sqrt{|\lambda_t|\tau_r}$ from sign schedule for current $\tau_m$
    \STATE Prepare HF, apply seeds and compiled orbital rotation, and measure $S$ occupation samples
  \ENDFOR
  \STATE Postselect raw outcomes to $(N_\alpha,N_\beta)$ (with optional spin-flip and bootstrap CI)
  \STATE Apply recovery policy, postselect again, and merge valid strings into cumulative count maps
  \STATE Solve selected CI on at most $K_{\mathrm{fb}}$ strings per spin sector
  \IF{$a>0$ and variational improvement $< \delta E_{\mathrm{stop}}$}
    \STATE break
  \ENDIF
\ENDFOR
\STATE Construct final product space according to $\mathcal F$ and solve the physical Hamiltonian
\RETURN $E_{\mathrm{var}}$, its lowest eigenvalue
\end{algorithmic}
\end{algorithm}

\subsection{Electronic Hamiltonian and variational output}

We consider an active space of $n$ spatial orbitals with fixed electron numbers $(N_\alpha,N_\beta)$ and Hamiltonian
\begin{equation}
\hat H=E_{\mathrm{nuc}}+\sum_{pq,\sigma}h_{pq}a^\dagger_{p\sigma}a_{q\sigma}
+\frac12\sum_{pqrs}\sum_{\sigma\tau}g_{pqrs}
a^\dagger_{p\sigma}a^\dagger_{q\tau}a_{s\tau}a_{r\sigma}.
\label{eq:hamiltonian}
\end{equation}
The quantum stage returns occupation samples, not an energy. Distinct retained alpha and beta strings define a determinant product space $\mathcal V$, and the uncorrected algorithmic output is
\begin{equation}
E_{\mathrm{var}}=\min_{0\ne|\psi\rangle\in\mathcal V}
\frac{\langle\psi|\hat H|\psi\rangle}{\langle\psi|\psi\rangle}.
\label{eq:variational-energy}
\end{equation}
Because the selected basis is orthonormal and Eq.~\eqref{eq:hamiltonian} is evaluated from the original active-space tensors, $E_{\mathrm{var}}$ is an upper bound to the exact ground-state energy of that active-space Hamiltonian. This statement does not apply to the PT2-corrected estimates introduced below.

\subsection{Natural-orbital preprocessing and correlated seeds}
\label{subsec:preprocessing}

The active-space integrals are rotated to an MP2 natural-orbital basis~\cite{jensen1988mp2no}, and the same rotated tensors are used in every classical solve. Coupled-cluster singles-and-doubles (CCSD) amplitudes~\cite{purvis1982ccsd} initialize a small number of number-preserving pair rotations but do not contribute an energy to Eq.~\eqref{eq:variational-energy}. A selected pair substitution $(i_\alpha i_\beta)\rightarrow(a_\alpha a_\beta)$ is encoded with parity-computing CNOTs and an $\mathrm{XX+YY}$ rotation. Later rounds replace the initial amplitudes with damped ratios extracted from the selected-CI state. Gate-level details are given in Supplementary Note S1.

\subsection{Double-factorized proposal directions}

Following low-rank and double-factorization constructions for electronic-structure Hamiltonians~\cite{motta2021lowrank}, the spatial electron-repulsion tensor is reshaped into an orbital-pair matrix and symmetrized before spectral decomposition,
\begin{equation}
g_{pqrs}=\sum_{t=1}^{n^2}\lambda_t A^{(t)}_{pq}A^{(t)}_{rs}.
\label{eq:df}
\end{equation}
The complete decomposition is an exact tensor identity. The circuit construction retains only a set $\mathcal L_L$ of $L$ leaves and diagonalizes each real symmetric leaf as $A^{(t)}=O_t\operatorname{diag}(d_t)O_t^{\mathsf T}$. Leaves are ranked by
\begin{equation}
s_t=|\lambda_t|\,\|A^{(t)}_{ov}\|_F^2,
\label{eq:leaf-score-methods}
\end{equation}
which combines factorization strength with occupied--virtual coupling relative to the reference. This is a support-oriented ranking, not the Frobenius-optimal rank-$L$ tensor truncation.

Define the spin-summed one-body operator
\begin{equation}
\hat B_t=\sum_{pq,\sigma}A^{(t)}_{pq}a^\dagger_{p\sigma}a_{q\sigma}.
\end{equation}
For a Hermitian $\hat B_t$, the Hubbard--Stratonovich Gaussian characteristic-function construction~\cite{hubbard1959partition,hirsch1983discrete} gives the exact single-factor identity
\begin{equation}
e^{-\tau|\lambda_t|\hat B_t^2/2}
=\mathbb E_z\!\left[e^{iz\sqrt{\tau|\lambda_t|}\hat B_t}\right].
\label{eq:hs-methods}
\end{equation}
Each field realization is a one-body orbital rotation. Products of all selected leaf rotations and the feedback Thouless rotation are multiplied classically at the one-particle level and compiled as one number-preserving orbital-rotation network. Equation~\eqref{eq:hs-methods} defines an auxiliary proposal factor. It is not an exact decoupling of the physical interaction used in Eq.~\eqref{eq:hamiltonian}: the circuit generator uses selected leaves, $|\lambda_t|$ and ordinary squares $\hat B_t^2$. The physical quartic interaction is associated with normal-ordered squares, whereas an ordinary square also contains a one-body contraction. No claim of physical imaginary-time propagation is made. Supplementary Notes S2 and S3 derive these distinctions and the ordered-product error.

\subsection{Deterministic auxiliary fields}

For circuit $r$ and leaf $t$, the implemented field is
\begin{equation}
\phi_t^{(r)}=h_{rt}\sqrt{|\lambda_t|\tau_r},\qquad h_{rt}\in\{-1,+1\},
\label{eq:hadamard-methods}
\end{equation}
where $h_{rt}$ is drawn from a Hadamard sign table. A complete balanced block of nonconstant columns has zero column means and diagonal covariance, exactly matching the first two moments of independent Gaussian fields at fixed $\tau$. The executed schedules use finite row and column subsets, include a constant Hadamard column in current implementations, and in some cases truncate the available row block. They are therefore deterministic sign schedules rather than guaranteed second-moment cubatures. Their empirical means and covariance defects must be reported per run rather than inferred from the ideal complete design. The construction removes randomness in choosing field directions; it does not remove quantum shot noise, hardware noise or errors due to unmatched higher moments.

The coherent average of the field unitaries can be compared perturbatively with a filter for
\begin{equation}
\hat V_{\mathrm{prop}}=\frac12\sum_{t\in\mathcal L_L}|\lambda_t|\hat B_t^2.
\end{equation}
The experiment instead measures each unitary separately and pools probabilities, eliminating cross-circuit interference. The coherent-filter analysis therefore motivates the proposal directions but is not a convergence theorem for the measured state and does not establish targeting of the physical ground state.

\subsection{Sampling, postselection and adaptive feedback}

Each circuit prepares the Hartree--Fock occupation, applies the selected pair seeds and then the compiled orbital rotation. Measurements are split into alpha and beta strings. In noiseless simulation, direct postselection retains only strings with Hamming weights $(N_\alpha,N_\beta)$. For hardware data, occupancy-guided configuration recovery~\cite{robledo2025sqd,qiskit_addon_sqd} can first propose corrections to number-violating outcomes; every proposal is then subjected to the same strict postselection. Recovery is a heuristic error-mitigation step and does not certify the pre-noise configuration. Spin-flipped partners are added in classical post-processing where specified.

During feedback, sampled weights are marginalized separately over alpha and beta strings. The top $K$ strings in each sector define a product space of dimension at most $K^2$. A selected-CI solve yields wavefunction marginals and excitation-amplitude ratios. Singles define a clipped Thouless rotation, while linked pair and generalized-double ratios define additional seed angles through $\theta=2\arctan r$. Updates are damped and skipped when the reference coefficient is too small. Ranking-only pseudo-counts derived from selected-CI marginals affect subsequent selection but do not enter Hamiltonian matrix elements. The reported runs used different frozen proposal settings. N$_2$ hardware used 16 leaves, 12 circuits, four pair seeds and no adaptive round. The Fe$_2$S$_2$ simulator point used 48 leaves and 24 circuits, while Fe$_2$S$_2$ hardware used 36 leaves and 24 circuits; all reported results unless otherwise stated are the initial \texttt{flip0} ensemble rather than adaptive-round outputs. The broader implementation supports selected-CI feedback, damping, Thouless clipping and linked double seeds as described above, but those features are not validated on hardware. Table~\ref{tab:hardware-proposal-parameters} gives the hardware settings used for these experiments.

\begin{table}[H]
\centering
\small
\begin{tabularx}{\linewidth}{@{}lXX@{}}
\toprule
\textbf{Parameter} & \textbf{N$_2$/6-31G hardware} & \textbf{Fe$_2$S$_2$ hardware} \\
\midrule
Executed ensemble & \texttt{flip0}; $A=0$ & \texttt{flip0}; $A=0$ \\
Leaves and fields $(L,R)$ & $(16,12)$ & $(36,24)$ \\
Shots & $S=8{,}192$; total $98{,}304$ & $S=16{,}384$; total $393{,}216$ \\
Time policy & $\eta=0.30$; $\boldsymbol\mu=(0.25,0.5,1.0)$ & $\eta=0.30$; $\boldsymbol\mu=(0.25,0.5,1.0)$ \\
Initial pair seeds & $P=4$ & $P=6$ \\
\bottomrule
\end{tabularx}
\caption{\textbf{Executed DF-SQD hardware proposal and sampling parameters.} The time scale is $\tau_0=\eta/\max(\chi,\varepsilon)$ with $\chi=\max_{t,i,a}|\lambda_t||A^{(t)}_{ia}|^2$, and $\tau_m=\mu_m\tau_0$.}
\label{tab:hardware-proposal-parameters}
\end{table}

\subsection{Support recovery and nested-space convergence}

Let the fixed executed circuit ensemble define a mixture $q(D)=\sum_rw_r|\langle D|U_r|\Phi_0\rangle|^2$. For independent mixture draws and a finite target set $\mathcal T$ with $q(D)\ge q_{\min}>0$, a union bound gives
\begin{equation}
\Pr(\mathcal T\nsubseteq\mathcal S_N)\le |\mathcal T|e^{-Nq_{\min}}.
\label{eq:support-bound-methods}
\end{equation}
The result is conditional on the executed mixture assigning nonzero probability to the target. For the fixed per-circuit allocation used experimentally, if circuit $r$ supplies $N_r$ independent shots and assigns probability $p_r(D)$ to $D$, then the miss probability is $\prod_r(1-p_r(D))^{N_r}\le\exp[-\sum_rN_rp_r(D)]$; a union bound gives the corresponding finite-target guarantee. Supplementary Note S5 explains both forms.

If cumulative sets of alpha and beta strings are never pruned, their Cartesian product spaces are nested and the corresponding Rayleigh--Ritz energies cannot increase. In the finite active-space Hilbert space, the sequence stabilizes almost surely at the product space generated by all accessible strings. Its limit equals the exact active-space energy only when that product space contains the exact ground-state wavefunction. Changing top-$K$ feedback spaces need not be nested and are not covered by this monotonicity statement.

\subsection{EN-inspired marginal-sector selection}

For the Fe$_2$S$_2$ hardware analysis, we developed a Hamiltonian-aware selection procedure tailored to spin-sector products. Starting from a normalized selected-CI coefficient array $c$ in a candidate alpha--beta product and its electronic variational energy $E_{\mathrm{elec}}=c^{\mathsf T}Hc$, each determinant pair $(i,j)$ receives the nonnegative score
\begin{equation}
s_{ij}^{\mathrm{EN}}=
\frac{|(Hc)_{ij}|^2}
{\max\!\left(|E_{\mathrm{elec}}-H_{ij,ij}|,\,0.02\hartree\right)}.
\label{eq:en-score}
\end{equation}
The denominator floor limits near-degeneracy sensitivity. Pair scores are marginalized over the opposite spin sector,
\begin{equation}
s_i^\alpha=\sum_j s_{ij}^{\mathrm{EN}},\qquad
s_j^\beta=\sum_i s_{ij}^{\mathrm{EN}},
\end{equation}
and the top $K$ strings from each ranking define a Cartesian product of at most $K^2$ determinants. We call this procedure \emph{EN-inspired marginal-sector selection}. It differs from conventional determinant-level CIPSI~\cite{huron1973cipsi} because it ranks spin sectors and retains their full product. The initial nominal 50-million-determinant construction began from a frequency-selected $K_{\mathrm{ref}}=2,000$ state and targeted $K=7,072$, corresponding to $50,013,184$ determinant pairs. Three fixed EN-score/select/selected-CI passes were run. During this construction, large boosts protected strings already represented in the current selected-CI support, so the update did not intentionally discard them.

The reverse $K$-sweep used a different, diagnostic protocol. One EN-inspired scoring pass from the final $K=7,072$ state produced fixed alpha and beta rankings. Every smaller $K$ used prefixes of those rankings, followed by an approximate selected-CI diagonalization; the $K=7,072$ state was reused. Scores were not recomputed at each $K$, and the sweep is therefore not an independent iterative CIPSI calculation at every point. Although the defined spaces are nested, the resulting iterative-solver energies contain nonmonotonic points, so the curve is not used as evidence for Rayleigh--Ritz monotonicity.

\subsection{Pool-restricted, denominator-clipped second-order correction}

For each selected product $\mathcal V_K=S_K^\alpha\times S_K^\beta$, the PT2 external space is the remainder of the finite post-configuration-recovery product pool,
\begin{equation}
\mathcal E_K=(U_\alpha\times U_\beta)\setminus\mathcal V_K.
\end{equation}
Using the selected-CI state $|\Psi_K\rangle$, the deterministic correction is
\begin{equation}
E^{(2)}=\sum_{D\in\mathcal E_K}
\frac{|\langle D|H|\Psi_K\rangle|^2}{\widetilde\Delta_D},\qquad
\widetilde\Delta_D=\min(E_{\mathrm{var}}^{\mathrm{elec}}-H_{DD},-0.02\hartree).
\label{eq:pt2}
\end{equation}
The clipping forces every denominator to be negative and no closer to zero than $-0.02\hartree$, thereby modifying the standard EN-PT2 estimator and guaranteeing a nonpositive correction by construction. This is a deterministic, sampled-pool-restricted, EN-inspired second-order estimate: the external space is not the full Hilbert space, no semistochastic completion is used, and the corrected energy is not re-diagonalized. PT2 was evaluated post hoc and did not affect quantum sampling or adaptive feedback.

For \method{} at $K=7,072$, $E_{\mathrm{var}}=-116.5654471004\hartree$ and $E^{(2)}=-0.0111408604\hartree$, giving errors of $40.162$ and $29.021\mha$ before and after PT2, respectively. For SQD-LUCJ, the corresponding values were $E_{\mathrm{var}}=-116.5338537174\hartree$ and $E^{(2)}=-0.0039576772\hartree$, with errors of $71.755$ and $67.798\mha$. The primary manuscript conclusions use the variational values; the clipped second-order estimate is reported only as a secondary diagnostic of Hamiltonian coupling within the recovered pool.

\subsection{Molecular systems and reference energies}

The N$_2$ hardware benchmark used an internuclear separation of $1.0977$~\AA{}, the 6-31G basis and a $(10e,16o)$ active space. Errors were computed relative to the CASSCF/FCI value $-109.10301625689416\hartree$. The comparator used an SQD-LUCJ proposal circuit.

The iron--sulfur benchmark used the $M_S=0$ sector of the Li--Chan active-space Hamiltonian for $[\mathrm{Fe}_2\mathrm{S}_2(\mathrm{SCH}_3)_4]^{2-}$ with 30 electrons in 20 spatial orbitals, $N_\alpha=N_\beta=15$, mapped to 40 qubits by the Jordan--Wigner transformation~\cite{jordan1928wigner,li_chan_fes_repository}. The selected product spaces do not enforce exact spin purity; the reported variational states have $\langle S^2\rangle\approx0.05$ and are compared with a singlet DMRG reference. Errors use the published $M=8,000$ density-matrix-renormalization-group reference $-116.6056091\hartree$. The calculations consume the active-space FCIDUMP directly; we therefore do not assign an atomic-orbital basis to that file without confirmation from its original provenance.

\subsection{Simulation and hardware execution}

Electronic-structure preprocessing used PySCF~\cite{sun2018pyscf}. Simulator circuits were compiled with optimization level=3 for FakeKingston as target and sampled using Qiskit Aer matrix-product-state simulation~\cite{qiskit2024} with maximum bond dimension of 256. Multiple transpiler seeds were evaluated and the lowest two-qubit-depth realization was retained.


Hardware circuits were transpiled to \texttt{ibm\_kingston} device and submitted. Counts, job identifiers, shot totals, transpiled depths and selected transpiler seeds were retained separately before pooling. The Fe$_2$S$_2$ comparison used one SQD-LUCJ circuit with 1,500,000 shots and 24 \method{} circuits with 16,384 shots each. The N$_2$ \method{} run used 12 circuits with 8,192 shots each. Reported timing panels retain the metrics recorded by each run; QPU usage is distinguished from simulator sampling wall time.






\section*{Data and materials availability}
All data supporting the findings is available in the main text and Supplementary materials. The github repository for codes to reproduce all results will be released following the conclusion of the review process.



\clearpage
\bibliographystyle{unsrt}
\bibliography{references}

\clearpage
\appendix
\renewcommand{\thesection}{S\arabic{section}}
\renewcommand{\thesubsection}{S\arabic{subsection}}
\renewcommand{\theequation}{S\arabic{equation}}
\renewcommand{\thefigure}{S\arabic{figure}}
\renewcommand{\thetable}{S\arabic{table}}
\renewcommand{\theHequation}{S\arabic{equation}}   
\renewcommand{\theHfigure}{S\arabic{figure}}       
\renewcommand{\theHtable}{S\arabic{table}}         
\setcounter{section}{0}
\setcounter{section}{0}
\setcounter{subsection}{0}
\setcounter{equation}{0}
\setcounter{figure}{0}
\setcounter{table}{0}
\setcounter{footnote}{0}                        
\renewcommand{\theHfootnote}{S\arabic{footnote}} 

\newcommand{\scititle}{DF-SQD: Deterministic Fields for Sampling-Based Quantum Diagonalization}

\renewcommand{\thefigure}{S\arabic{figure}}
\renewcommand{\thetable}{S\arabic{table}}
\renewcommand{\theequation}{S\arabic{equation}}

\begin{center}
{\Large\bfseries Supplementary Information for\\\scititle}\\[1em]
Kushagra Agarwal\footnote{Kushagra worked on this project during his summer internship at IBM Research} and Anupama Ray\\[0.35em]
\small IBM Research\\[0.35em]
\small Corresponding author: anupamar@in.ibm.com\\

\end{center}

\subsubsection*{Supplementary Note S1. Orbital preprocessing and correlated seeds}
\label{supp:preprocessing}

This note provides the mathematical details of natural-orbital preprocessing and correlated seeds. Let $U_{\mathrm{NO}}$ be the orthogonal
matrix whose columns are the MP2 natural orbitals~\cite{jensen1988mp2no}.  The one- and two-electron
integrals transform as
\begin{align}
 h^{\mathrm{NO}}_{ij}
 &=\sum_{pq}(U_{\mathrm{NO}})_{pi}h_{pq}(U_{\mathrm{NO}})_{qj},\\
 g^{\mathrm{NO}}_{ijkl}
 &=\sum_{pqrs}(U_{\mathrm{NO}})_{pi}(U_{\mathrm{NO}})_{qj}
   (U_{\mathrm{NO}})_{rk}(U_{\mathrm{NO}})_{sl}g_{pqrs}.
 \label{eq:supp-no-rotation}
\end{align}
The transformation is a basis change, not an approximation.  All subsequent
selected-CI calculations use the same tensors
$(h^{\mathrm{NO}},g^{\mathrm{NO}})$.

CCSD~\cite{purvis1982ccsd} is run in this basis to obtain an initial proposal, but its energy is not
used as the output of the algorithm.  Consider occupied spatial orbital $i$ and
virtual orbital $a$.  In the Jordan--Wigner qubit ordering used here, let
$q_{\alpha i},q_{\alpha a},q_{\beta i},q_{\beta a}$ denote their four occupation
qubits.  The implemented pair-seed block is
\begin{equation}
 U_{ia}(\theta)=
 \operatorname{CNOT}_{\alpha i\rightarrow\beta i}
 \operatorname{CNOT}_{\alpha a\rightarrow\beta a}
 \operatorname{XX{+}YY}_{\alpha i,\alpha a}(\theta,0)
 \operatorname{CNOT}_{\alpha i\rightarrow\beta i}
 \operatorname{CNOT}_{\alpha a\rightarrow\beta a},
 \label{eq:supp-pair-circuit}
\end{equation}
where operators act from right to left.  The outer CNOT pairs compute and then
uncompute the alpha--beta occupation parities.  Between them, the
$\mathrm{XX+YY}$ gate transfers one excitation between the alpha occupied and
virtual qubits.  On the paired reference pattern, the parity wiring makes this
transfer coincide with the beta transfer after uncomputation.  Restricted to
the two-dimensional pair subspace
$\operatorname{span}\{|\Phi_0\rangle,|\Phi_{ii}^{aa}\rangle\}$, the block acts,
up to a convention-dependent phase, as
\begin{align}
 U_{ia}(\theta)|\Phi_0\rangle
 &=\cos(\theta/2)|\Phi_0\rangle
   +e^{i\varphi_{ia}}\sin(\theta/2)|\Phi_{ii}^{aa}\rangle,\\
 U_{ia}(\theta)|\Phi_{ii}^{aa}\rangle
 &=\cos(\theta/2)|\Phi_{ii}^{aa}\rangle
   -e^{-i\varphi_{ia}}\sin(\theta/2)|\Phi_0\rangle.
 \label{eq:supp-pair-action}
\end{align}
The phase $\varphi_{ia}$ has no effect on occupation probabilities for an
isolated seed.  Hence
\begin{equation}
 P(\Phi_0)=\cos^2(\theta/2),
 \qquad
 P(\Phi_{ii}^{aa})=\sin^2(\theta/2),
 \qquad
 P(D\notin\{\Phi_0,\Phi_{ii}^{aa}\})=0.
 \label{eq:supp-pair-probabilities}
\end{equation}
This is the central reason for using the parity-entangled seed instead of an
unconditioned orbital transfer: in the ideal isolated block, every probability
moved out of the reference is assigned to the chemically selected paired double,
not to unrelated singles or wrong-particle-number strings.

If a desired pair probability is $p_{ia}$, Eq.~\eqref{eq:supp-pair-probabilities}
gives $\theta=2\arcsin\sqrt{p_{ia}}$.  The implementation treats the bounded
CCSD amplitude itself as the signed target sine amplitude and uses
\begin{equation}
 \theta_{ia}^{(0)}=2\arcsin\!\left[
 \operatorname{clip}(t_{ii}^{aa},-1,1)\right].
 \label{eq:supp-initial-angle}
\end{equation}
It follows immediately that the intended pair probability contributed by one
isolated seed is
\begin{equation}
 P_{ia}=|\operatorname{clip}(t_{ii}^{aa},-1,1)|^2.
 \label{eq:supp-amplitude-probability}
\end{equation}
Ranking seeds by $|t_{ii}^{aa}|$ therefore also ranks these ideal pair
probabilities.  If $|t_{ii}^{aa}|\ll1$, the reference probability changes only
by $\mathcal O(|t_{ii}^{aa}|^2)$, so small, chemically unimportant amplitudes do
not create a large diffuse background.

For several seeds, the gates need not commute when they share an occupied or
virtual orbital.  If their angles are small, expanding the ordered product gives
\begin{equation}
 \prod_{k=1}^{K_1}U_k(\theta_k)|\Phi_0\rangle
 =\left[1-\frac18\sum_k\theta_k^2+\mathcal O(\theta^3)\right]|\Phi_0\rangle
 +\sum_k e^{i\varphi_k}\frac{\theta_k}{2}|\Phi_k^{\mathrm{pair}}\rangle
 +\mathcal O(\theta^2).
 \label{eq:supp-multiple-seeds}
\end{equation}
Thus the selected pair doubles appear at first order in their angles and with
probability $\theta_k^2/4+\mathcal O(\theta^3)$, while determinants requiring
products of two different seed operations appear only at second order in
amplitude and fourth order in probability.  This is the precise perturbative
sense in which the seed layer concentrates probability on the chosen pair
doubles while suppressing combinatorial cross terms.  It is not a claim that
all other determinants remain exactly absent after multiple overlapping seeds,
the HS orbital rotation, or hardware noise.

On hardware, gate errors can create single substitutions, particle-number
violations, and other unintended outcomes.  Such leakage is not part of the
ideal probability statement.  Particle-number postselection and configuration
recovery described in Supplementary Note S7 remove or repair this leakage before
the classical solve.  Only finite seed amplitudes are retained, so a failed or
divergent CCSD amplitude cannot be converted into an invalid circuit angle.

\subsubsection*{Supplementary Note S2. Double factorization and leaf ranking}
\label{supp:double-factorization}

The spatial electron-repulsion tensor is matricized as
$G_{(pq),(rs)}=g^{\mathrm{NO}}_{pqrs}$ and symmetrized numerically,
\begin{equation}
 G\leftarrow\frac12(G+G^{\mathsf T}).
 \label{eq:supp-supermatrix}
\end{equation}
Its complete spectral decomposition is the exact finite-dimensional identity
\begin{equation}
 G_{(pq),(rs)}
 =\sum_{t=1}^{n^2}\lambda_tA^{(t)}_{pq}A^{(t)}_{rs},
 \qquad A^{(t)}=A^{(t)\mathsf T}.
 \label{eq:supp-first-factorization}
\end{equation}
If only a selected set $\mathcal L_L$ of $L$ leaves is retained, the algorithm
uses
\begin{equation}
 G^{(L)}_{(pq),(rs)}
 =\sum_{t\in\mathcal L_L}\lambda_tA^{(t)}_{pq}A^{(t)}_{rs},
 \qquad R_L=G-G^{(L)}.
 \label{eq:supp-truncated-factorization}
\end{equation}
The residual $R_L$ is the leaf-truncation error.  For conventional spectral
truncation by $|\lambda_t|$, its Frobenius norm is the square root of the sum of
squared discarded eigenvalues.  The present method instead selects leaves by a
support-generation score, so it deliberately optimizes circuit relevance rather
than the globally minimal Frobenius reconstruction error.  Full factorization
and the second diagonalization below are exact; only replacing $G$ by $G^{(L)}$
is approximate.
Every leaf is diagonalized again,
\begin{equation}
 A^{(t)}=O_t\operatorname{diag}(d_t)O_t^{\mathsf T}.
 \label{eq:supp-second-factorization}
\end{equation}
Define the spin-summed one-body operator
\begin{equation}
 \hat B_t=\sum_{pq,\sigma}A^{(t)}_{pq}
 a^\dagger_{p\sigma}a_{q\sigma}.
 \label{eq:supp-leaf-operator}
\end{equation}
The normal-ordered square $:\!\hat B_t^2\!:$ generates the quartic interaction
associated with the factorized tensor.  The circuit construction instead uses
the ordinary square $\hat B_t^2$, which additionally contains a one-body
contraction through the anticommutation relation
$a_qa_r^\dagger=\delta_{qr}-a_r^\dagger a_q$; its explicit form depends on the
integral convention.  Consequently, the auxiliary circuit generator should not
be identified with the retained normally ordered physical interaction.  This
mismatch does not alter the Hamiltonian matrix elements used in the reported
selected-CI energy, because those matrix elements are evaluated from the
original active-space tensors.  It can nevertheless change which determinants
are proposed and thereby contribute indirectly to finite-support variational
error.

We now justify the occupied--virtual score.  Partition a leaf matrix into
occupied and virtual blocks relative to $|\Phi_0\rangle$.  For a small parameter
$\epsilon$,
\begin{equation}
 e^{i\epsilon\hat B_t}|\Phi_0\rangle
 =|\Phi_0\rangle+i\epsilon\hat B_t|\Phi_0\rangle
  +\mathcal O(\epsilon^2).
 \label{eq:supp-leaf-expansion}
\end{equation}
The occupied--occupied part merely changes the basis spanning the occupied
subspace, and the virtual--virtual part annihilates the reference at first order.
The occupied--virtual block generates substitutions with amplitudes proportional
to $A^{(t)}_{ia}$.  At the next order, products
$\lambda_tA^{(t)}_{ia}A^{(t)}_{jb}$ generate double-substitution structure.
Consequently,
\begin{equation}
 s_t=|\lambda_t|\|A^{(t)}_{ov}\|_F^2
 \label{eq:supp-leaf-score}
\end{equation}
measures both interaction strength and the capacity to move probability out of
the reference.  In particular, $A^{(t)}_{ov}=0$ makes the leaf inert as a
first-order support generator even if $|\lambda_t|$ is large.

For time-scale selection, define
\begin{equation}
 \chi=\max_{t\le L}\max_{i\in o,a\in v}
 |\lambda_t||A^{(t)}_{ia}|^2,
 \qquad
 \tau_0=\frac{\eta}{\max(\chi,\varepsilon)},
 \qquad
 \tau_m=\mu_m\tau_0.
 \label{eq:supp-time-scale}
\end{equation}
The parameter $\eta$ controls the largest estimated first-order double scale,
while several $\mu_m$ values diversify the sampled support.

\subsubsection*{Supplementary Note S3. Hubbard--Stratonovich identity and product error}
\label{supp:hs}

Using the Hubbard--Stratonovich construction~\cite{hubbard1959partition,hirsch1983discrete}, for a Hermitian operator $\hat B$ with spectral decomposition
$\hat B=\sum_bb\Pi_b$, the Gaussian characteristic-function identity gives
\begin{align}
 \mathbb E_{z\sim\mathcal N(0,1)}
 \left[e^{iz\sqrt{\tau\lambda}\hat B}\right]
 &=\sum_b\mathbb E_z[e^{iz\sqrt{\tau\lambda}b}]\Pi_b\\
 &=\sum_be^{-\tau\lambda b^2/2}\Pi_b\\
 &=e^{-\tau\lambda\hat B^2/2},
 \label{eq:supp-hs-proof}
\end{align}
for $\lambda\ge0$.  Thus HS decoupling is exact for a single positive
quadratic leaf.

For independent fields $z_t$, sequential expectation also proves the exact
ordered-product identity
\begin{equation}
 \mathbb E_{\boldsymbol z}\!\left[
 \prod_{t=1}^{L}e^{iz_t\sqrt{\tau|\lambda_t|}\hat B_t}
 \right]
 =\prod_{t=1}^{L}e^{-\tau|\lambda_t|\hat B_t^2/2}.
 \label{eq:supp-ordered-hs}
\end{equation}
No commutativity assumption is needed in Eq.~\eqref{eq:supp-ordered-hs}, because
integrating $z_t$ replaces the corresponding factor while preserving its
position in the ordered product.

Noncommutativity enters when this product is compared with the exponential of
the auxiliary proposal generator
\begin{equation}
 \hat V_{\mathrm{prop}}=\frac12\sum_{t=1}^{L}|\lambda_t|\hat B_t^2.
 \label{eq:supp-retained-generator}
\end{equation}
This notation is deliberately distinct from the retained physical interaction:
it uses absolute eigenvalues and ordinary squares.
Let $X_t=-\tau|\lambda_t|\hat B_t^2/2$.  The Lie product formula gives
\begin{equation}
 \prod_te^{X_t}=e^{-\tau\hat V_{\mathrm{prop}}}+\mathcal O(\tau^2).
 \label{eq:supp-lie-product}
\end{equation}
A finite-dimensional operator-norm estimate is
\begin{equation}
 \left\|\prod_te^{X_t}-e^{-\tau\hat V_{\mathrm{prop}}}\right\|
 \le \frac{\tau^2}{8}
 e^{\tau\sum_t|\lambda_t|\|\hat B_t\|^2/2}
 \sum_{t<u}|\lambda_t\lambda_u|
 \|[\hat B_t^2,\hat B_u^2]\|.
 \label{eq:supp-trotter-bound}
\end{equation}
The bound makes the approximation structure transparent: Gaussian decoupling is
exact for each stated ordinary-square auxiliary factor, whereas combining
noncommuting factors into a single exponential of $\hat V_{\mathrm{prop}}$ has a
local error quadratic in the short time.  This statement concerns the proposal
generator, not the physical interaction.  A symmetric ordering would improve
the product-formula order, but the frozen implementation uses a first-order
ordered product.

The implementation uses $|\lambda_t|$ in every field magnitude.  For
nonnegative retained modes this agrees with the sign entering the tensor
factorization, although the ordinary-square contraction distinction remains.
A materially negative mode changes the HS contour and requires a real-field
continuation; replacing $\lambda_t$ by $|\lambda_t|$ instead changes the
generator.  Thus, regardless of spectral checks, the implemented HS circuits
are support proposals derived from the factorization, not an exact auxiliary-
field representation of the physical Hamiltonian.

Finally, diagonalizing $A^{(t)}$ gives the one-particle rotation
\begin{equation}
 W_r^{\mathrm{HS}}
 =\prod_{t=1}^{L}
 O_t\operatorname{diag}(e^{i\phi_t^{(r)}d_t})O_t^{\mathsf T}.
 \label{eq:supp-field-rotation}
\end{equation}
Every factor is unitary.  Their matrix product is computed classically and its
second-quantized action is compiled as one number-preserving orbital rotation.

\subsubsection*{Supplementary Note S4. Hadamard cubature and filter consistency}
\label{supp:hadamard}

Let $H_Q\in\{\pm1\}^{Q\times Q}$ be a Sylvester Hadamard matrix with
$Q\ge\max(L,R)$ a power of two.  At fixed time $\tau$, define
\begin{equation}
 \phi_t^{(r)}=h_{rt}\sqrt{|\lambda_t|\tau}.
 \label{eq:supp-hadamard-fields}
\end{equation}
For a complete balanced block of nonconstant columns,
\begin{align}
 \frac1Q\sum_rh_{rt}&=0,\\
 \frac1Q\sum_rh_{rt}h_{ru}&=\delta_{tu},\\
 \frac1Q\sum_r\phi_t^{(r)}\phi_u^{(r)}
 &=\tau|\lambda_t|\delta_{tu}.
 \label{eq:supp-hadamard-moments}
\end{align}
Thus the first two moments equal those of independent Gaussian fields with
variances $\tau|\lambda_t|$.

Define the ordered field unitary and its Hadamard average by
\begin{equation}
 U(\boldsymbol\phi)=\prod_te^{i\phi_t\hat B_t},
 \qquad
 \overline U_H(\tau)=\frac1Q\sum_rU(\boldsymbol\phi^{(r)}),
 \label{eq:supp-hadamard-average}
\end{equation}
and define the ideal Gaussian average
\begin{equation}
 U_G(\tau)=\mathbb E_{\boldsymbol z}
 \left[U(\sqrt{\tau|\boldsymbol\lambda|}\odot\boldsymbol z)\right]
 =\prod_te^{-\tau|\lambda_t|\hat B_t^2/2}.
 \label{eq:supp-gaussian-average}
\end{equation}
Expanding the ordered exponentials shows why moment matching is useful.  The
zero first moment cancels terms linear in $\sqrt\tau$, and covariance
orthogonality cancels unwanted mixed quadratic terms.  The first unmatched
contributions arise from third and higher moments, yielding
\begin{equation}
 \|\overline U_H(\tau)-U_G(\tau)\|
 \le C_3\tau^{3/2}+C_4\tau^2.
 \label{eq:supp-hadamard-error}
\end{equation}
If $b_t=\sqrt{|\lambda_t|}\|\hat B_t\|$ and
$M^{(3)}_{tuv}=Q^{-1}\sum_rh_{rt}h_{ru}h_{rv}$, the leading cubic contribution
can be bounded by
\begin{equation}
 C_3\le\frac16\sum_{tuv}|M^{(3)}_{tuv}|b_tb_ub_v.
 \label{eq:supp-third-constant}
\end{equation}
Hadamard column orthogonality alone does not imply that all third mixed moments
vanish.  If the design is centrally symmetric, with every $h_r$ paired with
$-h_r$ at the same $\tau$, all odd moments vanish and
\begin{equation}
 \|\overline U_H(\tau)-U_G(\tau)\|\le C_4\tau^2.
 \label{eq:supp-paired-error}
\end{equation}
Combining Eqs.~\eqref{eq:supp-trotter-bound} and
\eqref{eq:supp-hadamard-error} gives
\begin{equation}
 \|\overline U_H(\tau)-e^{-\tau\hat V_{\mathrm{prop}}}\|
 \le C_3\tau^{3/2}+(C_4+C_{\mathrm{LT}})\tau^2.
 \label{eq:supp-filter-consistency}
\end{equation}
This is a short-time coherent-filter consistency result for the auxiliary
proposal generator.

The frozen implementations use finite row and column subsets, include a constant
Hadamard column in the current schedules, and cycle several time values.  Some
runs also truncate the available balanced row block.  Exact zero mean and
diagonal covariance must therefore not be assumed for an executed schedule.  Its leading
moment defects are directly measurable:
\begin{align}
 m_t&=R^{-1}\sum_r\phi_t^{(r)},\\
 C_{tu}&=R^{-1}\sum_r\phi_t^{(r)}\phi_u^{(r)},\\
 \Delta_C&=C-\operatorname{diag}
 \left(R^{-1}\sum_r|\lambda_t|\tau_r\right).
 \label{eq:supp-empirical-defects}
\end{align}
Nonzero $m$ restores a first-order field term, while $\Delta_C$ quantifies the
second-moment mismatch.  Reporting these quantities is therefore a direct check
of how closely a finite circuit schedule realizes the ideal balanced rule.

There is also a conceptual distinction between the operator identity and the
experiment.  The coherent filtered amplitude uses
$\overline U_H|\Phi\rangle$.  Separate circuit measurement instead produces
\begin{equation}
 q_H(D)=\frac1Q\sum_r|\langle D|U_r|\Phi\rangle|^2.
 \label{eq:supp-mixture-probability}
\end{equation}
In general,
$|\langle D|Q^{-1}\sum_rU_r|\Phi\rangle|^2\ne q_H(D)$ because the latter lacks
cross-circuit interference.  Equation~\eqref{eq:supp-filter-consistency} therefore characterizes the
coherent average of the Hamiltonian-derived proposal circuits.  It does not
establish that the pooled distribution is an imaginary-time state or that it
preferentially targets the physical ground state.  The next note establishes
support recovery and variational-space convergence for the measured mixture.

\subsubsection*{Supplementary Note S5. Sampling, variational convergence, and imaginary time}
\label{supp:convergence}

\paragraph{Theorem S1 (finite-shot support recovery).}
Index all executed circuits by $r$ and let
\begin{equation}
 q(D)=\sum_rw_r|\langle D|U_r|\Phi_0\rangle|^2,
 \qquad w_r\ge0,\quad\sum_rw_r=1,
 \label{eq:supp-exact-mixture}
\end{equation}
be the probability of determinant $D$ under the shot-allocation mixture.  Let
$\mathcal T$ be a finite target set with $q(D)\ge q_{\min}>0$ for all
$D\in\mathcal T$.  If $\mathcal S_N$ is the support observed in $N$ independent
shots, then
\begin{equation}
 \Pr[\mathcal T\nsubseteq\mathcal S_N]
 \le |\mathcal T|e^{-Nq_{\min}}.
 \label{eq:supp-coverage-bound}
\end{equation}

\paragraph{Proof.}
For one determinant, the probability of being missed in every shot is
$(1-q(D))^N$.  Since $1-x\le e^{-x}$,
\begin{equation}
 (1-q(D))^N\le e^{-Nq(D)}\le e^{-Nq_{\min}}.
\end{equation}
A union bound over the $|\mathcal T|$ possible missed determinants proves
Eq.~\eqref{eq:supp-coverage-bound}.  For a fixed per-circuit allocation with
$N_r$ independent shots from circuit $r$ and
$p_r(D)=|\langle D|U_r|\Phi_0\rangle|^2$, the corresponding miss probability is
\begin{equation}
 \Pr[D\ \mathrm{missed}]
 =\prod_r(1-p_r(D))^{N_r}
 \le \exp\!\left[-\sum_rN_rp_r(D)\right].
 \label{eq:supp-stratified-bound}
\end{equation}
A union bound over a finite target set gives the stratified analogue used for
fixed experimental shot allocations.  For the mixture-draw form, therefore
\begin{equation}
 N\ge q_{\min}^{-1}\log(|\mathcal T|/\zeta)
 \label{eq:supp-shot-complexity}
\end{equation}
is sufficient for recovery probability at least $1-\zeta$.
\hfill$\square$

For a fixed finite circuit ensemble and indefinitely continued independent
sampling, the theorem implies that every determinant of nonzero mixture
probability is eventually observed almost surely.  It does not assume that the
circuit mixture equals a coherent imaginary-time state.

\paragraph{Theorem S2 (monotone convergence of cumulative uncapped spaces).}
Let $\mathcal A_N$ and $\mathcal B_N$ be the cumulative sets of symmetry-valid
alpha and beta half-strings observed after $N$ shots, including classical
spin-flip augmentation, and define
\begin{equation}
 \mathcal V_N=\operatorname{span}
 \{|I_\alpha J_\beta\rangle:I\in\mathcal A_N,J\in\mathcal B_N\}.
 \label{eq:supp-cumulative-space}
\end{equation}
Let $E_N$ be the lowest Rayleigh quotient of $\hat H$ in $\mathcal V_N$, and
let $E_{\mathrm{gs}}$ denote the exact active-space ground-state energy.  Then
\begin{equation}
 \mathcal V_N\subseteq\mathcal V_{N+1},
 \qquad E_{\mathrm{gs}}\le E_{N+1}\le E_N.
 \label{eq:supp-monotone-energy}
\end{equation}
For a fixed finite circuit ensemble,
\begin{equation}
 E_N\xrightarrow[N\rightarrow\infty]{\mathrm{a.s.}}E_\infty,
 \label{eq:supp-accessible-limit}
\end{equation}
where $E_\infty$ is the minimum in the product space generated by all
half-strings of nonzero marginal probability.

\paragraph{Proof.}
Cumulative sampling never removes an observed half-string.  Both half-string
sets and therefore their Cartesian-product spaces are nested.  Minimizing a
Rayleigh quotient over an enlarged space cannot increase its value, proving
Eq.~\eqref{eq:supp-monotone-energy}.  The active-space Hilbert space is finite.
By Theorem S1, each positive-probability half-string is eventually observed
almost surely, while a zero-probability half-string is never observed.  The
space therefore stabilizes almost surely at the accessible product space,
proving Eq.~\eqref{eq:supp-accessible-limit}.
\hfill$\square$

\paragraph{Corollary S2.1.}
If the accessible product space contains the exact ground state, then
$E_\infty=E_{\mathrm{gs}}$.  Otherwise, $E_\infty-E_{\mathrm{gs}}$ is the residual variational error due
to circuit-inaccessible support.  A top-$K$ feedback space can replace strings
between rounds and need not be nested; the theorem therefore applies to the
cumulative uncapped construction and final solve.

We next state the ideal imaginary-time result used to motivate the circuit
family and clarify the limited
coherent-filter interpretation of the circuit construction.  Let a Hermitian
generator $\hat G$ have eigenvalues
$G_0<G_1\le\cdots\le G_{\max}$ and gap $\Delta_G=G_1-G_0$.  Expand the initial
state as $|\Phi\rangle=\sum_jc_j|\Psi_j^G\rangle$, with $c_0\ne0$, and define
\begin{equation}
 |\Phi(T)\rangle=
 \frac{e^{-T\hat G}|\Phi\rangle}
      {\|e^{-T\hat G}|\Phi\rangle\|}.
 \label{eq:supp-normalized-ite}
\end{equation}
Relative to the filtered ground component, the squared magnitude of excited
component $j$ is
$|c_j/c_0|^2e^{-2T(G_j-G_0)}$.  Summing these terms gives
\begin{equation}
 1-|\langle\Psi_0^G|\Phi(T)\rangle|^2
 \le\frac{1-|c_0|^2}{|c_0|^2}e^{-2T\Delta_G},
 \label{eq:supp-ite-overlap}
\end{equation}
and multiplication by the spectral width gives
\begin{equation}
 0\le\langle\Phi(T)|\hat G|\Phi(T)\rangle-G_0
 \le(G_{\max}-G_0)
 \frac{1-|c_0|^2}{|c_0|^2}e^{-2T\Delta_G}.
 \label{eq:supp-ite-energy}
\end{equation}
Thus ideal imaginary time converges exponentially for nonzero initial overlap
and a nonzero gap.

Suppose an approximate short step satisfies
\begin{equation}
 \|F_\tau-e^{-\tau\hat G}\|\le c\tau^{1+\rho},
 \qquad\rho>0.
 \label{eq:supp-local-filter-error}
\end{equation}
For $M=T/\tau$, insert and subtract one exact factor at a time:
\begin{equation}
 F_\tau^M-e^{-T\hat G}
 =\sum_{k=0}^{M-1}F_\tau^{M-1-k}
   (F_\tau-e^{-\tau\hat G})e^{-k\tau\hat G}.
 \label{eq:supp-telescoping}
\end{equation}
Taking norms yields
\begin{equation}
 \|F_\tau^M-e^{-T\hat G}\|
 \le\mathcal O(T\tau^\rho)e^{T\|\hat G\|}.
 \label{eq:supp-global-filter-error}
\end{equation}
A centrally symmetric Hadamard rule has local second-order error and hence
first-order global convergence at fixed $T$.  A second-moment-only rule can
retain a local $\tau^{3/2}$ term, giving global order $\tau^{1/2}$.

These imaginary-time results apply only to coherent composition of the averaged
filter and, in the present construction, only to the auxiliary proposal
generator $\hat V_{\mathrm{prop}}$.  That generator uses selected leaves,
absolute factorization eigenvalues, and ordinary squares, and it omits the
remaining terms of the physical Hamiltonian.  The implemented experiment
additionally pools measurements from separate one-step unitaries rather than
coherently composing averaged filters.  Therefore, these equations do not imply
convergence toward the physical ground state and do not constitute a convergence
result for the measured protocol.  Theorems S1 and S2 provide the applicable
guarantees: recovery of nonzero-probability support and monotone Rayleigh--Ritz
convergence in cumulative uncapped spaces.  The HS--Hadamard analysis supplies a
controlled coherent-filter interpretation for the proposal generator only.

\subsubsection*{Supplementary Note S6. Marginal selection and adaptive feedback}
\label{supp:feedback}

Let $p(I,J)$ be the normalized postselected weight of a determinant with alpha
half-string $I$ and beta half-string $J$.  Define
\begin{equation}
 p_\alpha(I)=\sum_Jp(I,J),
 \qquad
 p_\beta(J)=\sum_Ip(I,J).
 \label{eq:supp-sample-marginals}
\end{equation}
Let $\mathcal A_K$ and $\mathcal B_K$ contain the $K$ largest corresponding
marginals.  The feedback product space obeys
\begin{equation}
 \mathcal V_K=\operatorname{span}
 \{|I_\alpha J_\beta\rangle:I\in\mathcal A_K,J\in\mathcal B_K\},
 \qquad\dim\mathcal V_K\le K^2.
 \label{eq:supp-sector-space}
\end{equation}

The marginal rule is optimal within each spin sector.  For any alpha set $A$
of cardinality at most $K$,
\begin{equation}
 \sum_{I\in A}p_\alpha(I)
 \le\sum_{I\in\mathcal A_K}p_\alpha(I).
 \label{eq:supp-topk-optimality}
\end{equation}
To prove this, suppose $A$ contains a marginal smaller than one omitted from
$A$.  Exchanging the two cannot decrease retained mass.  Repeating the exchange
produces $\mathcal A_K$.  The beta proof is identical.  This proves separate
marginal optimality, not global optimality over arbitrary sets of $K^2$ joint
determinants.

If $C=(c_{IJ})$ is the selected-CI coefficient matrix, the wave-function
marginals are
\begin{equation}
 w_\alpha(I)=\sum_J|c_{IJ}|^2,
 \qquad
 w_\beta(J)=\sum_I|c_{IJ}|^2.
 \label{eq:supp-sci-marginals}
\end{equation}
The leading marginal strings are encoded as large pseudo-counts and merged into
the ranking pool.  They influence only which half-strings survive the next cap;
they are not quantum frequencies and do not enter Hamiltonian matrix elements.

For feedback, let $c_{\mathrm{HF}}$ be the current reference coefficient.  The
spin-averaged single ratio is
\begin{equation}
 r_i^a=\frac{c_{I(i\rightarrow a),J_{\mathrm{HF}}}
                 +c_{I_{\mathrm{HF}},J(i\rightarrow a)}}
                {2c_{\mathrm{HF}}}.
 \label{eq:supp-single-ratio}
\end{equation}
The linked pair and generalized-double ratios are
\begin{align}
 r_{ii}^{aa,\mathrm{link}}
 &=\frac{c_{I(i\rightarrow a),J(i\rightarrow a)}}{c_{\mathrm{HF}}}
   -(r_i^a)^2,\\
 r_{ij}^{ab,\mathrm{link}}
 &=\frac{c_{I(i\rightarrow a),J(j\rightarrow b)}}{c_{\mathrm{HF}}}
   -r_i^ar_j^b.
 \label{eq:supp-linked-ratios}
\end{align}
The subtraction removes disconnected products of singles already generated by
the orbital rotation.

With damping $\gamma$,
\begin{equation}
 r^{(a+1)}=(1-\gamma)r^{(a)}+\gamma r^{\mathrm{SCI}}.
 \label{eq:supp-damped-update}
\end{equation}
The singles define
\begin{equation}
 \kappa_i^a=\operatorname{clip}
 (\arctan r_i^a,-\kappa_{\max},\kappa_{\max}),
 \qquad U_{\mathrm{relax}}=e^{K(\kappa)},
 \label{eq:supp-thouless-update}
\end{equation}
and each retained linked ratio becomes a two-level angle
\begin{equation}
 \theta=2\arctan r.
 \label{eq:supp-feedback-angle}
\end{equation}
The $2\arctan$ relation follows from an amplitude ratio
$r=\tan(\theta/2)$.  Since every ratio is divided by $c_{\mathrm{HF}}$, feedback
becomes ill-conditioned if that coefficient vanishes.  The implementation
therefore skips updates below a numerical threshold, damps every update, and
clips the Thouless elements.  These safeguards prevent the formal ratio map
from turning a small reference coefficient into uncontrolled circuit angles.

\begin{figure}[H]
\centering
\includegraphics[width=0.8\textwidth]{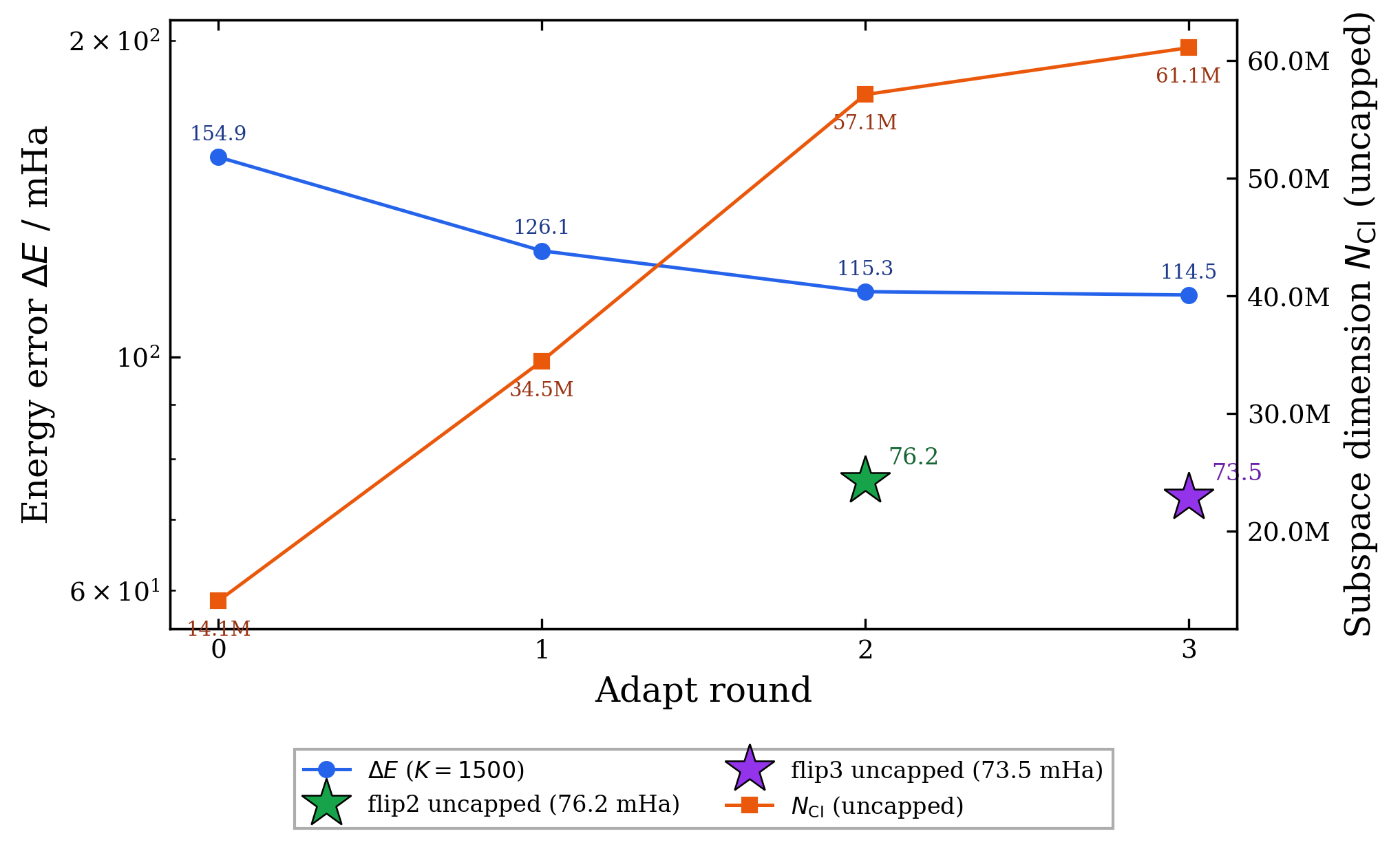}
\caption{\textbf{Adaptive feedback and classical-space growth for the Fe$_2$S$_2$ simulator baseline.} Variational selected-CI error relative to the Li--Chan $M=8{,}000$ DMRG reference, $E_{\mathrm{ref}}=-116.6056091$~Ha, is shown across the initial ensemble (\texttt{flip0}) and three cumulative adaptive ensembles (\texttt{flip1}--\texttt{flip3}). Blue circles use the feedback cap $K_{\mathrm{fb}}=1{,}500$ strings per spin sector ($N_{\mathrm{CI}}=2.25\times10^6$); stars show separate uncapped solves of the \texttt{flip2} and \texttt{flip3} cumulative pools. Orange squares give the available uncapped alpha--beta Cartesian-product dimension at each checkpoint; round-0 and round-1 uncapped energies were not evaluated. Each round used 24 FakeKingston-transpiled Aer-MPS circuits with 8,192 shots per circuit. Checkpoints after \texttt{flip0} are cumulative and include selected-CI ranking pseudo-counts used for adaptive feedback. ``Uncapped'' denotes all strictly postselected, spin-flip-augmented strings in the finite cumulative pool, not the full-CI space.}
\label{fig:supp-baseline-adapt}
\end{figure}

Figure~\ref{fig:supp-baseline-adapt} evaluates the classical-space growth and variational-energy improvement for a simulated adaptive Fe$_2$S$_2$ run over three rounds. The exact stored run configuration for this adaptive-baseline evaluation is recorded in Table~\ref{tab:supp-baseline-adapt}. Note that this configuration uses $8$ double-factorization leaves, differing from the $48$-leaf and $36$-leaf parameters of the main-text single-round Fe$_2$S$_2$ results.

\begin{table}[H]
\centering
\small
\begin{tabularx}{\linewidth}{@{}lX@{}}
\toprule
\textbf{Parameter} & \textbf{Value} \\
\midrule
Retained leaves & 8 \\
Field circuits & 24 \\
Time multipliers & 0.25, 0.5, 1.0 \\
Amplitude target & 0.3 \\
Initial pair / adaptive generalized seeds & 6 / 3 \\
Thouless cap & 0.5 \\
Shots per circuit & 8,192 \\
Adaptive rounds & 3 \\
Adaptive damping & 0.6 \\
Adaptive tolerance & 0.0 mHa \\
Feedback cap $K$ & 1,500 (used for the capped $\Delta E$ line) \\
SQD configuration-recovery iterations & 1 \\
\bottomrule
\end{tabularx}
\caption{\textbf{Configuration parameters for the Fe$_2$S$_2$ adaptive-baseline evaluation plotted in Fig.~\ref{fig:supp-baseline-adapt}.}}
\label{tab:supp-baseline-adapt}
\end{table}

\paragraph{Why the Thouless update recenters the sampler.}
Let $\kappa$ be an occupied--virtual matrix and define the anti-Hermitian
single-particle generator
\begin{equation}
 K(\kappa)=
 \begin{pmatrix}
  0 & \kappa\\
  -\kappa^{\mathsf T} & 0
 \end{pmatrix},
 \qquad U_{\mathrm{relax}}=e^{K(\kappa)}.
 \label{eq:supp-thouless-matrix}
\end{equation}
Its second-quantized action is
\begin{equation}
 \widehat U_{\mathrm{relax}}
 =\exp\!\left[
 \sum_{ia,\sigma}\kappa_i^a
 (a^\dagger_{a\sigma}a_{i\sigma}
  -a^\dagger_{i\sigma}a_{a\sigma})
 \right].
 \label{eq:supp-thouless-operator}
\end{equation}
Thouless' theorem~\cite{thouless1960stability} states that any Slater determinant nonorthogonal to
$|\Phi_0\rangle$ can be represented in this form.  Its first-order expansion is
\begin{equation}
 \widehat U_{\mathrm{relax}}|\Phi_0\rangle
 =|\Phi_0\rangle+
 \sum_{ia,\sigma}\kappa_i^a|\Phi_{i\sigma}^{a\sigma}\rangle
 +\mathcal O(\kappa^2).
 \label{eq:supp-thouless-expansion}
\end{equation}
The measured single ratios therefore estimate the occupied--virtual tangent
direction from the old reference toward the best determinant represented by
the current selected-CI state.  Multiplying
$W_r^{\mathrm{HS}}U_{\mathrm{relax}}$ makes every next-round field fluctuate
around this rotated occupied subspace.  Since the product of one-particle
unitaries is again a one-particle unitary, the relaxation can be folded into the
same orbital-rotation circuit rather than appended as a separate network.

\subsubsection*{Supplementary Note S7. Configuration recovery, strict postselection, and spin-flip augmentation}
\label{supp:configuration-recovery}

Ideal circuit components preserve $N_\alpha$ and $N_\beta$, but hardware errors
can change one or both measured Hamming weights.  Direct postselection is always
valid, but it discards every such shot.  Configuration recovery instead treats
a particle-number-violating sample as a noisy observation of an unknown valid
configuration and proposes a corrected string before strict postselection.

Let $x_{p\sigma}\in\{0,1\}$ be a measured occupation and let
$n_{p\sigma}\in[0,1]$ be the current average orbital occupancy obtained from the
selected-CI solution.  If a spin half-string contains too many electrons,
occupied bits are candidates for removal, with orbitals of small
$n_{p\sigma}$ favored.  If it contains too few electrons, empty bits are
candidates for insertion, with orbitals of large $n_{p\sigma}$ favored.  The
procedure repeats independently in the two spin sectors until the requested
Hamming weights are reached.  Schematically, the correction weights are
\begin{align}
 P(p\text{ removed}\mid x_{p\sigma}=1)&\propto1-n_{p\sigma},\\
 P(p\text{ inserted}\mid x_{p\sigma}=0)&\propto n_{p\sigma}.
 \label{eq:supp-recovery-weights}
\end{align}
When several noisy strings recover to the same determinant, their weights are
combined and renormalized.

Recovery does not certify that the corrected determinant was the exact
pre-noise configuration, and it is not part of the coherent-filter analysis or
the support-recovery theorem.  Its purpose is error mitigation: it recycles
shots using physically informed
occupancies instead of assigning all number-violating outcomes zero weight.
Every recovered string is subsequently subjected to strict
$(N_\alpha,N_\beta)$ postselection, so an invalid particle-number sector can
never enter selected CI.  For ideal or noiseless simulation data, where the
circuits already preserve particle number, recovery makes no useful change and
is omitted; direct postselection is used instead.

After recovery and postselection, spin-flip augmentation maps each count key
$(J_\beta,I_\alpha)$ to $(I_\alpha,J_\beta)$ with the same multiplicity.  This
adds the partner obtained by exchanging the spin halves and makes the sampled
support more compatible with a spin-symmetric classical solve.  It is a
classical support symmetrization and should not be confused with adding spin-
dependent auxiliary-field circuits.

\begin{figure}[H]
\centering
\subfloat[N$_2$ DF-SQD, circuit 5.]{\includegraphics[width=0.47\textwidth]{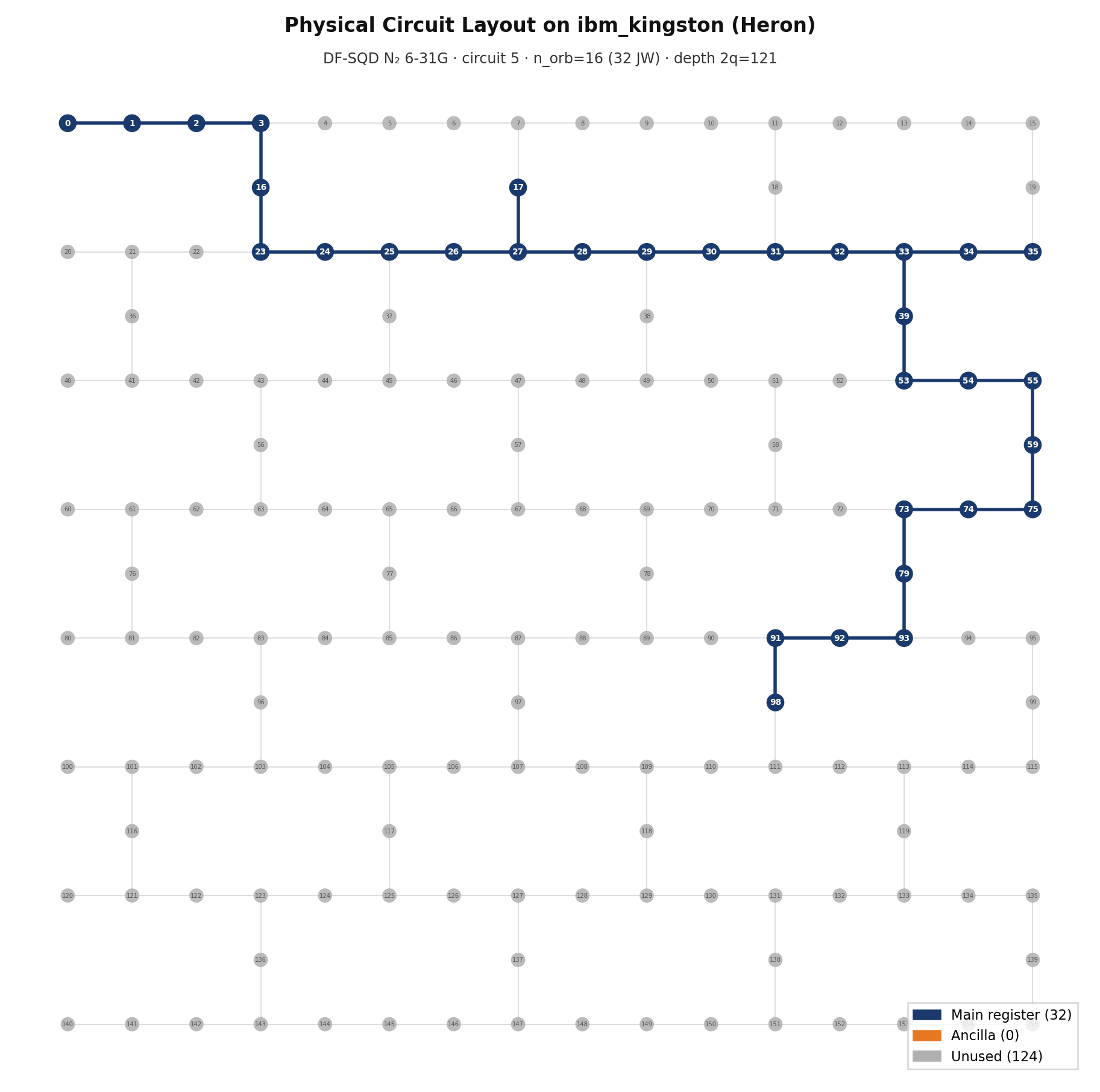}}\hfill
\subfloat[Fe$_2$S$_2$ DF-SQD, circuit 1.]{\includegraphics[width=0.47\textwidth]{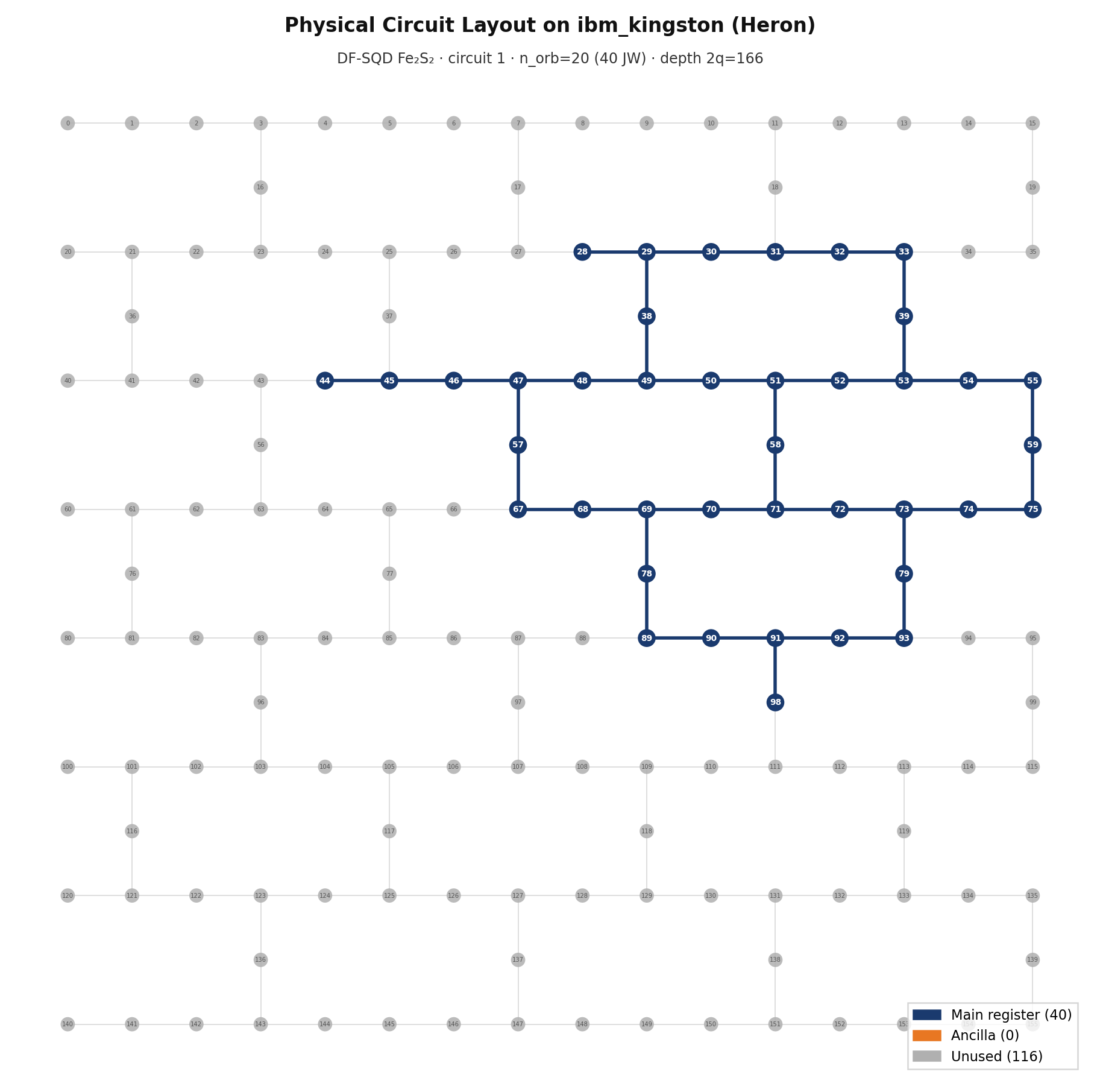}}
\caption{\textbf{Representative DF-SQD placements on \texttt{ibm\_kingston}.} These are transpiler-selected placements on the 156-qubit device, not fixed chemistry mappings or evidence of an algorithm-independent routing advantage.} 
\label{fig:kingston-layouts}
\end{figure}

Representative transpiler-selected \method{} placements on the 156-qubit \texttt{ibm\_kingston} device are shown in Fig.~\ref{fig:kingston-layouts}. The N$_2$ example is circuit 5 (2q-depth 121; 32 main and no ancilla qubits; 8,192 shots in a 12-circuit batch), and the Fe$_2$S$_2$ example is circuit 1 (2q-depth 166; 40 main and no ancilla qubits; 16,384 shots in a 24-circuit job).

\end{document}